\documentclass[longbib,PRB,reprint,superscriptaddress]{revtex4-1}
\usepackage{amsmath}
\usepackage{graphicx}
\usepackage{braket}
\usepackage{url}
\usepackage[T1]{fontenc}
\usepackage[utf8]{inputenc}
\usepackage{amssymb}
\usepackage{longtable}
\usepackage[unicode=true]{hyperref}
\usepackage[usenames,dvipsnames]{color}

\begin{abstract}
We investigate the many-body properties of graphene on top of a piezoelectric substrate, focusing on the interaction between the graphene electrons and the piezoelectric acoustic phonons. We calculate the electron and phonon self-energies as well as the electron mobility limited by the substrate phonons. We emphasize the importance of the proper screening of the electron-phonon vertex and discuss the various limiting behaviors as a function of electron energy, temperature, and doping level. The effect on the graphene electrons of the piezoelectric acoustic phonons is compared with that of the intrinsic deformation acoustic phonons of graphene.
Substrate phonons tend to dominate over intrinsic ones for low doping levels at high and low temperatures.
\end{abstract}

\makeatletter

\begin{document}

\title{Many-body effects in doped graphene on a piezoelectric substrate}
\author{David G. Gonz\'alez}
\email[email:\,]{d.gonzalez@mat.ucm.es}
\affiliation{Departamento de F\'isica de Materiales, Universidad Complutense de Madrid, E-28040 Madrid, Spain}
\affiliation{Campus de Excelencia Internacional, Campus Moncloa UCM-UPM, E-28049 Madrid, Spain}
\author{Ivar Zapata}
%\email[email:\,]{izapatao@ucm.es}
\affiliation{Departamento de F\'isica de Materiales, Universidad Complutense de Madrid, E-28040 Madrid, Spain}
\author{J\"urgen Schiefele}
\affiliation{Departamento de F\'isica de Materiales, Universidad Complutense de Madrid, E-28040 Madrid, Spain}
\affiliation{%
Instituto de Ciencia de Materiales de Madrid, CSIC, E-28\,049 Madrid, Spain}
\author{Fernando Sols}
\affiliation{Departamento de F\'isica de Materiales, Universidad Complutense de Madrid, E-28040 Madrid, Spain}
\affiliation{Campus de Excelencia Internacional, Campus Moncloa UCM-UPM, E-28049 Madrid, Spain}
\author{Francisco Guinea}
\affiliation{%
IMDEA Nanociencia, Calle de Faraday~9,
E-28\,049 Madrid, Spain}
\affiliation{%
Department of Physics and Astronomy, University of Manchester, Oxford Road, Manchester M13~9PL, United Kingdom
}

\date{\today}
%\currfilename
\maketitle

\section{Introduction}

Elastic waves supported by the boundaries of solids
and, in particular, surface acoustic waves (SAWs),
underlie numerous applications of microwave devices for signal processing \cite{Weigel_2002}.
SAWs with amplitudes of a few nanometers can be electrically excited on the surface of piezoelectric materials,
and the resulting periodic deformation in adjacent thin film materials or quantum-well structures
can be employed
to modulate optical resonances  in polaritonic or plasmonic devices \cite{Cerda-Mendez_2013,Ruppert_2010,Schiefele2013}.
Apart from the mechanical deformation, the vibration of the ionic lattice in a piezoelectric material  produces an
electric field travelling along with the SAW,
which can transport charge carriers in monolayer graphene deposited on top of the piezomaterial \cite{Santos_2013,Miseikis2012,Bandhu2013}
and, for instance, probe graphene's Landau level structure when an external magnetic field is applied \cite{Thalmeier2010}.

Because the carbon allotrope graphene is an atomically thin all-surface material \cite{CastroNeto2009},
its charge carrier dynamics is very sensitive to the surrounding electromagnetic fields,
and the possibility of changing graphene's carrier concentration \emph{in situ} by applying an external gate voltage
is a key feature in many graphene-based devices \cite{Ferrari_2014}.
Ballistic charge transport in suspended graphene over micrometer distances
and unprecedented carrier mobilities \cite{Bolotin_2008} are enabled by the high frequencies of the
optical phonons in the stiff honeycomb lattice. Thus,
the effects of electron-phonon scattering on transport are small in comparison with  conventional metals
\cite{CastroNeto2009}.
However, in most device architectures, graphene is deposited on a substrate,
and all lattice modes of the substrate material that induce an electric field will influence the carriers in the graphene
sheet, making the choice of substrate material crucial for the resulting transport characteristics of the device \cite{Dean_2010}.
This mechanism of remote phonon scattering in graphene has been mainly studied for substrates supporting optical phonon modes
\cite{Fratini2008,Schiefele_2012,Amorim_2012b}.

In the present work, we aim to clarify the role of acoustic piezoelectric surface phonons, which form the microscopic quanta of SAWs \cite{Ezawa_1971},
in graphene-on-piezomaterial structures.
After analyzing within a diagrammatic framework the effective carrier interaction due to exchange of surface phonons in Sec.~\ref{subsec:ep_eff},  we study the self-energies acquired by both phonons and
charge carriers in Sections~\ref{sec:phonon_SE} and \ref{sec:electron_SE}.
While the renormalization of the Fermi velocity due to piezoelectric substrate phonons turns out to be small,
we show that there are regimes where the substrate effects dominate momentum relaxation mechanism in graphene.
% \cite{Zhang2013}.
We compare both lifetimes and mobilities with the results obtained when only intrinsic acoustic deformation phonons are considered.
The numerical results for mean free paths and electron mobilities shown in Sec.~\ref{sec:numerical} are applicable to a variety of piezoelectrical
materials with different lattice structures and piezoelectric strengths.
Our study can be relevant for graphene devices operating in the ballistic transport regime such as hot electron transistor devices \cite{Tse2008} or field effect transistors based on graphene on different piezoelectrics \cite{Hong2009,Bidmeshkipour2015}, and for scenarios where quantum interference induces localization phenomena \cite{Li2013}.

%for which we list the relevant material parameters.

%\begin{itemize}
%\item We focus on the imaginary and real parts of the on-shell electron self-energies. The former shows interesting results that can be relevant for graphene devices operating in the ballistic transport regime such as hot electron transistor devices \cite{Tse2008}, and affect some quantum interference induced localization phenomena \cite{Li2013}. The latter have influence on the Fermi velocity and their effects turn out to be smaller.
%
%\end{itemize}

\section{Effective electron-electron interaction}
\label{subsec:ep_eff}
%%%%%%%%%%%%%%%%%%%%%%%%%%%%%%%%%%%%%%%%%%%%%%%%%%%%%%%%%%%%%%%%%%%%%%%%%%%%%%%%%%%%%%%%%%%%%%%%%%%%
%In the following, we describe the dynamics of charge carriers in monolayer graphene deposited
%on a piezoelectric substrate material with the Hamiltonian
%\begin{eqnarray}
%H= \sum_{\mathbf{k},\sigma}E_{\mathbf{k}} a_{\mathbf{k},\sigma}^{\dagger} a_{\mathbf{k},\sigma} + \hbar \sum_{\mathbf{q}}\omega_{\mathbf{q}}\, b_{\mathbf{q}}^{\dagger} b_{\mathbf{q}} \nonumber \\
%+ \frac{1}{2A} \sum_{\bf{q}} v_{\mathbf{q}}^{(0)} \rho(\mathbf{q}) \rho(\mathbf{-q} )~
%+  H_{\text{e-ph}}^{\text{PA}}
%.
%\label{eqn:sum_of_H}
%\end{eqnarray}
%The first two terms describe the the systems of charge carriers  and phonons seperately,
%$E_{\mathbf{k}}$ and $\hbar\omega_{\mathbf q}$ denoting their respective unperturbed dispersion relations.
%
%The piezoelectric substrate influences the carrier dynamics in graphene via the abovementioned piezoelectric
%electron-phonon coupling, as well as by modifying the screening between carriers.
%This electron-phonon interaction is valid for frequencies in the range of the
%sound ones, so that the optical phonons are integrated out by the use of the
%mentionend static dielectric constant.
%
The sound velocities $v_s(\theta)$ of piezoelectric acoustic phonons are anisotropic (depend on direction angle $\theta$) and typically two or three orders of magnitude smaller than
the Fermi velocity $v_F$ in graphene, which yields a relatively low value of the maximum acoustic frequency.
%In fact, in the derivation of any piezoelectric electron-phonon
%interaction, it is taken into account that the Fr\"ohlich optical phonons
%respond instantaneously in the time scales of the acoustic phonons by
%implicitly taking the effective dielectric constants of the problem as the
%static ones.
Thus the dielectric screening effects due to the substrate can be described by its static (also anisotropic) dielectric constant $\varepsilon_{0}(\theta)$.
This constant combines both core excitons and optical phonons as
well as any high-frequency (instantaneous) polarization forces which screen the fields created by the piezoelectric acoustic phonons \cite{Mahan2013}. 
The Fourier transform of the repulsive Coulomb interaction thus reads %(see Appendix \ref{app:opticalPhonons})
\begin{equation}
v_{\mathbf{q}}^{(0)}=\frac{2\pi e^{2}}{\overline{\varepsilon}_{0}(\theta) q}~,
\label{eqn:v_0}
\end{equation}
where $\overline{\varepsilon}_{0}(\theta) = \frac{\varepsilon_{0}(\theta)+1}{2}$ is the
effective dielectric constant at the substrate--air interface \cite{Simon1996}, and $q=|\mathbf{q}|$ with $\mathbf{q}=(q_x,q_y)$
and $\theta\equiv \arg(q_x+iq_y)$.
%:::::::::::::::::::::::::::::::::::::::::::::::::

%::::::::begin 2a
The interaction between the graphene electrons and the piezoelectric acoustic (PA) phonons is given by
\begin{align}
	H_{\text{e-ph}}^{\text{PA}}=\frac{1}{\sqrt{A}} {\sum_{\mathbf{k,q},\sigma}}\gamma_{\mathbf{q}}^{\text{PA}} \, a_{\mathbf{k+q},\sigma}^{\dagger}a_{\mathbf{k},\sigma}b_{\mathbf{q}}+\text{H.c.} 
	\label{eqn:PAHamiltonian}
\end{align}
Here, $A$ is the sample area, $a_{\mathbf{k},\sigma}$ is the Fermi operator for 
an electron of wave vector $\hbar \mathbf{k}$, spin-valley-cone index $\sigma$, and 
energy
\begin{equation}
E_{k\sigma}=\hbar s v_F k \, ,
\end{equation} 
where $s=\pm 1$ is the cone index; $b_{\mathbf{q}}$
is the Bose operator for a substrate PA phonon of wave vector $\mathbf{q}$ and (direction dependent) frequency
\begin{equation}
	\omega_{\mathbf{q}}=v_s(\theta)q \, .
\end{equation}
As in Ref. \cite{Gonzalez2015}, we assume that we only have to deal with phonons of momentum much smaller than the piezoelectric inverse lattice spacing (elastic limit). The electron-phonon coupling is then characterized by the $q$-independent vertex
%:::::::::end 3a

%\begin{equation}
%\gamma^{\text{PA}}_\mathbf{q} = K_R(\theta) \sqrt{\frac{\pi \alpha_\text{fs}\hbar^2 v_Fv_s(\theta)}{\overline{\varepsilon}_{0}(\theta)}} ~,
%\label{eqn:PA_vertex_def}
%\end{equation}
\begin{equation}
\gamma^{\text{PA}}_\mathbf{q} = K_R(\theta)
\left [\frac{\pi \alpha_\text{fs}\hbar^2 v_Fv_s(\theta)}{\overline{\varepsilon}_{0}(\theta)}  \right]^{\frac{1}{2}}
 ~,
\label{eqn:PA_vertex_def}
\end{equation}
where
$\alpha_\text{fs}=e^2/(\hbar v_F) \simeq 2.2$.
It is to be understood that, in the absence of the substrate, we are in the usual Fermi liquid regime of (doped) graphene.
Quite generally, the electromechanical coupling
coefficient satisfies $K_R(\theta) <1$.
The general derivation of Eq.~(\ref{eqn:PA_vertex_def}) is discussed in Ref.~\cite{Gonzalez2015}.

The coupling Eq.~(\ref{eqn:PA_vertex_def}) enables a
%the bare (that is, unscreened)
phonon-mediated electron-electron interaction
%\begin{equation}
%V_{\text{ph}}^{\text{PA}}(\mathbf{q},\omega)
%	= |\gamma_{\mathbf{q}}^{\text{PA}}|^{2} G_{0}^{\text{PA}}(\mathbf{q},\omega)
%	= 2|\gamma_{\mathbf{q}}^{\text{PA}}|^{2} \frac{\hbar v_{s}q}{\hbar^{2}\omega^{2}-\hbar^{2}v_{s}^{2}q^{2}} ~,
%\end{equation}
%\begin{eqnarray}
%V_{\text{ph}}^{\text{PA}}(\mathbf{q},\omega)
%&=& |\gamma_{\mathbf{q}}^{\text{PA}}|^{2} G_{0}^{\text{PA}}(\mathbf{q},\omega) \nonumber \\
%&=& |\gamma_{\mathbf{q}}^{\text{PA}}|^{2} \frac{2 v_{s}q/\hbar}{\omega^{2}-v_{s}^{2}q^{2}} ~,
%\end{eqnarray}
\begin{equation}
	V_{\text{ph}}^{\text{PA}}(\mathbf{q},\omega)
	= |\gamma_{\mathbf{q}}^{\text{PA}}|^{2} G_{0}^{\text{PA}}(\mathbf{q},\omega)
	\label{eqn:V-eff-PA-q-omega}
%	\nonumber \\
%	&=& |\gamma_{\mathbf{q}}^{\text{PA}}|^{2} \frac{2 v_{s}q/\hbar}{\omega^{2}-v_{s}^{2}q^{2}}
~,
\end{equation}
where
\begin{equation}
G_{0}^{\text{PA}}(\mathbf{q},\omega)=\frac{2\omega_\mathbf{q}/\hbar}{\omega^2-\omega_\mathbf{q}^2+i0^+}
\label{eqn:bare-propagator}
\end{equation}
denotes the bare propagator of the surface acoustic phonons.
By including screening effects due to  the charge carriers in graphene, as described by the polarization $\Pi_0(q,\omega)$ (see Refs.~\cite{Wunsch2006,Hwang2007}),
we can define the total effective electron-electron interaction in terms of an anisotropic dielectric function
$\varepsilon(\mathbf{q},\omega)$:
%The dielectric function for the effective interaction is derived in an analogous fashion (Eq. \ref{eqn:V_effOpticalPhonon}):
%\begin{equation}
%V_{\text{eff}}(\mathbf{q},\omega)=\frac{2\pi e^2}{\varepsilon(\mathbf{q},\omega)q}=\frac{v_{\mathbf{q}}^{(0)}+V_{\text{ph}}^{\text{PA}}(\mathbf{q},\omega)}{1-\left(v_{\mathbf{q}}^{(0)}+V_{\text{ph}}^{\text{PA}}(\mathbf{q},\omega)\right)\,\Pi_{0}(q,\omega)} ~.
%\label{eqn:V_eff}
%\end{equation}
\begin{eqnarray}
V_{\text{eff}}(\mathbf{q},\omega)
&=& \frac{2\pi  e^2}{\varepsilon(\mathbf{q},\omega)q} \nonumber \\
&=& \frac{v_{\mathbf{q}}^{(0)}+
	V_{\text{ph}}^{\text{PA}}(\mathbf{q},\omega)}{1-\left[ v_{\mathbf{q}}^{(0)}+V_{\text{ph}}^{\text{PA}}(\mathbf{q},\omega)\right]\,\Pi_{0}(q,\omega)} ~.
\label{eqn:V_eff}
\end{eqnarray}
Here, and in the following, we adopt the convention of referring to $q$ as a subindex or argument when the dependence on $\mathbf{q}$ has circular symmetry.
For low enough frequencies, typically $\hbar \omega \ll k_B T_{\rm BG}$ where $T_{\rm BG}$ is the Bloch-Gr\"uneisen temperature [defined in Eq. (\ref{eqn:TBG_def})], the phonon-induced electron-electron interaction adopts a $q$-dependence similar to that of Coulomb interaction:
\begin{equation}
V_{\text{ph}}^{\text{PA}}(\mathbf{q},\omega)
%=|\gamma_{\mathbf{q}}^{\text{PA}}|^{2}G_{0}^{\text{op}}(\mathbf{q},\omega \simeq 0) =
\simeq - \frac{2|\gamma_{\mathbf{q}}^{\text{PA}}|^{2}}{\hbar v_{s}q}~. \label{eqn:Vstatic}
\end{equation}
%It is, in this case, the acoustic phonon propagator $G_{0}^{\text{PA}}(\mathbf{q},\omega \simeq 0)$ that introduces the coulombic long-range dependence in $q$ via the dispersion of the modes, rather than the vertex.

By defining
%\begin{align}
%\varepsilon_{\text{RPA}}(\mathbf{q},\omega) &= 1- v_{\mathbf{q}}^{(0)}(\mathbf{q})\Pi_{0}(q,\omega) \label{eqn:dielectricFunctionRPA} \\
%& \left( \stackrel{\omega \rightarrow 0}{\longrightarrow}1+\frac{4 \alpha_\text{fs}k_F}{\overline{\varepsilon}_0\,q}=:1+\frac{k_\text{TF}}{q}\right)~, \label{eqn:dielectricFunctionRPAStaticLimit}
%\end{align}
\begin{equation}
	\varepsilon_{\text{RPA}}(\mathbf{q},\omega) = 1- v_{\mathbf{q}}^{(0)}\Pi_{0}(q,\omega) \, , \label{eqn:dielectricFunctionRPA}
\end{equation}
we obtain for $\omega \rightarrow 0$
\begin{equation}
\varepsilon_{\text{RPA}}(\mathbf{q},\omega)\simeq \varepsilon_{\text{RPA}}(\mathbf{q},0) \, ,
\end{equation}
where the static dielectric function satisfies,	
\begin{equation}
\varepsilon_{\text{RPA}}(\mathbf{q},0) =
%	 \stackrel{\omega \rightarrow 0}{\longrightarrow}
%	1+\frac{4 \alpha_\text{fs}k_F}{\overline{\varepsilon}_0\,q}=:
	1+\frac{k_\text{TF}(\theta)}{q}
	\label{eqn:dielectricFunctionRPAStaticLimit}
\end{equation}
for $q < 2k_F$, where 
%$k_{\rm TF}=4\alpha_{\rm fs}k_F/\bar{\varepsilon}_0(\theta)$
\begin{equation}
k_{\rm TF}=\frac{4\alpha_{\rm fs}k_F}{\bar{\varepsilon}_0(\theta)}
\end{equation}
is the (anisotropic) Thomas-Fermi wave vector and $k_F$ the Fermi wave vector, the factor of 4 accounting for spin-valley degeneracy.

We may also define the renormalized phonon propagator
\begin{align}
\tilde{G}^{\text{PA}}(\mathbf{q},\omega)&= \frac{G_{0}^{\text{PA}}(\mathbf{q},\omega)}
{1-\frac{
%|\gamma_{\mathbf{q}}^{\text{PA}}|^{2}
%		G_{0}^{\text{PA}}(\mathbf{q},\omega)
V_{\text{ph}}^{\text{PA}}(\mathbf{q},\omega)		
\Pi_{0}(q,\omega)}{\varepsilon_{\text{RPA}}(\mathbf{q},\omega)}}~.
\label{eqn:dressedPhononPropagator}
\end{align}
Then, Eq. (\ref{eqn:V_eff}) can be decomposed into an electron-electron and an
electron-phonon part \cite{Mahan2013,Mattuck2012}. We obtain \cite{Gonzalez2015}
\begin{figure}
	\centering
	\includegraphics[width=0.9\linewidth]{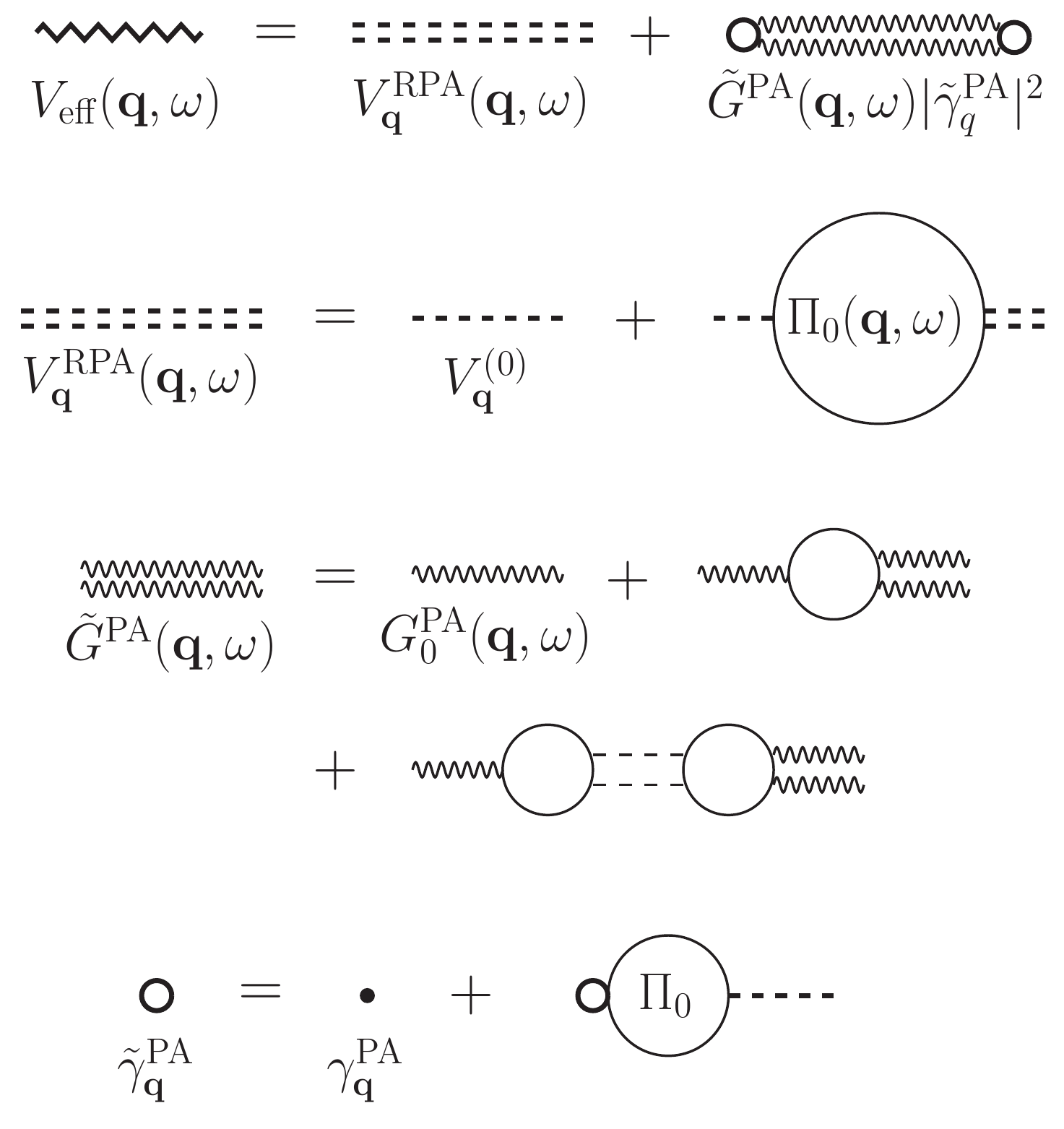}
	\caption{Equivalent RPA scheme for the effective electron-electron interaction
		separated into an electron-electron Coulombic part and an electron-phonon part
		with a screened vertex and renormalized phonon propagator. In the second line, $V^{\text{RPA}}(\mathbf{q},\omega)$ is the first term of Eq. (\ref{eqn:V_effAlternative})
	}
	\label{fig:VEffRPA_PhononElectronContributions}
\end{figure}
\begin{align}
 V_{\text{eff}}(\mathbf{q},\omega)
 	%&=\frac{v_{\mathbf{q}}}{\varepsilon(\mathbf{q},\omega)}  \\
	& = \frac{v_{\mathbf{q}}^{(0)}}{\varepsilon_{\text{RPA}}(\mathbf{q},\omega)}
	+ \left|\frac{\gamma_{\mathbf{q}}^{\text{PA}}}{\varepsilon_{\text{RPA}}(\mathbf{q},\omega)} \right|^{2} \tilde{G}^{\text{PA}}(\mathbf{q},\omega)~,
\label{eqn:V_effAlternative}
\end{align}
as shown diagrammatically in Fig.~\ref{fig:VEffRPA_PhononElectronContributions}.
We wish to emphasize the importance of electronic screening of the electron-phonon vertex shown in Eq. (\ref{eqn:V_effAlternative}). This will strongly influence the role of scattering processes involving low values of $q$.

A dimensionless parameter $\lambda_{\text{e-ph}}(\theta)$ characterizing the strength of the coupling of Eqs.~(\ref{eqn:PA_vertex_def}),(\ref{eqn:Vstatic})
can be obtained from multiplying the resulting effective interaction
(\ref{eqn:V-eff-PA-q-omega}) at $q=k_F$ by the density of states at the Fermi energy,
\begin{equation}
D(E_{F})
	%= \frac{2E_{F}}{\pi(\hbar v_{F})^{2}}
	= -\Pi_0(k_F,0)
	=\frac{2k_{F}}{\pi \hbar v_{F}}
	\, ,
\end{equation}
which leads to
%\begin{align}
%%V_{\text{ph}}^{\text{PA}}(\mathbf{q},0) \Pi_{0}(\mathbf{q},0)\sim \frac{4|\gamma_{\mathbf{q}}^{\text{PA}}|^{2}}{\pi \hbar^{2} v_{F} v_{s}} \equiv
%\lambda_{\text{e-ph}} (\theta)
%	&\equiv
%	V_{\text{ph}}^{\text{PA}}(k_F\hat{\mathbf{q}},0) \Pi_{0}(k_F,0)
%	=K_R^2(\theta)\frac{4\alpha_\text{fs}}{\overline{\varepsilon}_{0}(\theta)} \nonumber \\
%	&= 4 K_R^2 (\theta)r_s(\theta)
%,
%\label{eqn:lambdaDefinition}
%\end{align}
\begin{align}
	\lambda_{\text{e-ph}} (\theta)
	&\equiv
	V_{\text{ph}}^{\text{PA}}(k_F\hat{\mathbf{q}},0) \Pi_{0}(k_F,0)
%	K_R^2(\theta)\frac{4\alpha_\text{fs}}{\overline{\varepsilon}_{0}(\theta)}
		 \nonumber \\
	&=\frac{4}{\pi \hbar^2 v_s v_F }
	|\gamma_{\mathbf{q}}^{\text{PA}}|^{2}= 4 K_R^2 (\theta)r_s(\theta)
	,
	\label{eqn:lambdaDefinition}
\end{align}
where $\hat{\mathbf{q}}=\mathbf{q}/q$ and
the parameter
\begin{align}
r_s(\theta)\equiv\frac{\alpha_\text{fs}}{\overline{\varepsilon}_0(\theta)}
\label{eqn:rs_def}
\end{align}
characterizes the ratio between the interaction and kinetic energies.
This yields for the
ratio between the piezoelectric interaction and the residual static Coulomb
repulsion \cite{Gonzalez2015}
\begin{equation}
\frac{\lambda_{\text{e-ph}}}{\lambda_{\text{e-e}}} = K_R^{2}
\, ,
\label{eqn:ratio_e_p}
\end{equation}
%(the same parameter for the Coulomb repulsion would be $\lambda_\text{e-e}=4r_s$ and for the Fr\"ohlich optical phonons $\lambda_\text{e-op}=4\alpha_\text{fs}({\overline{\varepsilon}_\infty}^{-1}-{\overline{\varepsilon}_0}^{-1})$, see Appendix \ref{app:opticalPhonons}).
where
\begin{equation}
\lambda_{\rm e-e}(\theta)=v_{k_F\hat{\mathbf{q}}}^{(0)} \Pi (k_F,0)=\frac{4\alpha_\text{fs}}{\overline{\varepsilon}_0(\theta)}
=4r_s(\theta)
\end{equation}
is the dimensionless electron-electron coupling strength in substrate-screened graphene.

The electromechanical coupling coefficient $K_R(\theta)$, characteristic of each piezoelectric material,
can be measured in SAW experiments. It depends on the material's
piezoelectric, elastic and dielectric tensors, as well as on its mass density.
In Table \ref{tab:materials} we summarize angle-averaged values for selected representative materials as taken from Refs. \cite{Royer2000,Auld1990,Knabchen1996}.

\begin{table*}[htb]
	\begin{center}
		\begin{tabular}{|c|c|c|c|c|c|c|c|}
			\hline
			Material & Cut & $K_R^{2}$ & $\overline{\varepsilon}_0$ & $v_{s}$($\frac{\text{cm}}{\text{s}}$)& $|\gamma_{\mathbf{q}}^{\text{PA}}|^{2}$ $(\text{eV}\text{cm})^2$& $\lambda_\text{e-ph}$ & $4r_s$ \\
			\hline \hline
			GaAs (cubic) & X-Y-Z & $0.0015$ & 6.9 & $2.70\times10^{5}$ & $1.71\times10^{-20}$ & 0.0019 & 1.3 \\ \hline
			ZnO (6mm)& Z-Cut & $0.016$ & 4.8 & $2.71\times10^5$ & $2.70 \times 10^{-19}$ & $0.029$ & 1.8
			\\ \hline
			ZnO (6mm)& X-Cut & 0.0064 & 4.8 & $2.63\times10^5$ & $6.60\times10^{-20}$ & 0.0074 & 1.8 \\ \hline
			AlN (6mm)& Z-Cut & $0.0026$ & 5.0 & $5.85\times10^5$ & $9.18 \times 10^{-20}$ & $0.0046$ & 1.8 \\ \hline
			AlN (6mm)& X-Cut & $0.0048$ & 5.0 & $5.81\times10^5$ & $1.66\times10^{-19}$ & 0.0084 & 1.8 \\ \hline
			$\text{LiNbO}_{3}$ (3m) & Z-Cut & 0.0068 & 19 & $3.85\times10^5$ & $4.25\times10^{-20}$ & 0.0032 & 0.46 \\ \hline
			$\text{LiNbO}_{3}$ (3m) & Y-Cut & 0.017 & 20 & $3.59\times10^5$ & $9.35 \times 10^{-20}$ & 0.0077 & 0.44 \\ \hline
			$\text{LiNbO}_{3}$ (3m) & X-Cut & 0.019 & 20 & $3.60\times10^5$ & $9.80\times10^{-20}$ & 0.0080 & 0.44 \\ \hline
			PZT-4 (6mm)& Z-Cut & $0.027$ & 350 & $2.26\times10^5$ & $5.37\times10^{-21}$ & $7.0\times10^{-4}$ & 0.025 \\ \hline
			PZT-4 (6mm)& X-Cut & 0.0021 & 350 & $1.80\times10^5$ & $3.17\times10^{-22}$ & $5.2\times10^{-5}$ & 0.025 \\ \hline
		\end{tabular}
		\caption{%
			Angle-averaged values of the electromechanical coupling coefficient
			$K_R^2$
			appearing in Eq.~(\ref{eqn:PA_vertex_def}), the effective dielectric constant $\bar{\varepsilon_0}$, the sound velocity $v_s$, the vertex strength $\gamma_{\mathbf{q}}^{\text{PA}}$ [see Eq.~(\ref{eqn:PA_vertex_def})], the dimensionless coupling strength $\lambda_{\text{e-ph}}$ defined in  Eq.~(\ref{eqn:lambdaDefinition}), and the ratio $k_{\rm TF}/k_F$
			for several materials. Numerical values of the elastic tensors have been taken from Refs.~\cite{Royer2000,Auld1990,Knabchen1996} and references therein.}
		\label{tab:materials}
	\end{center}
\end{table*}

For example, the materials considered in Ref. \cite{Schiefele2013}, namely, \text{ZnO} and \text{AlN}, have associated piezoelectric tensors that are much larger than those of \text{GaAs} \cite{Pedros2011}, which increases the electron-phonon coupling by more than one order of magnitude. But there exist piezoelectric materials whose coefficients are even larger, like e.g.
$\text{LiNbO}_{3}$, $\text{BaTiO}_{3}$ or the PZT (lead zirconate titanate) $\text{PbTi}_{x}\text{Zr}_{1-x}\text{O}_{3}$, among many oxides with the perovskite structure and formula $\text{ABO}_{3}$, which tend to show ferroelectric properties, and are sometimes reminiscent of the layers between $\text{CuO}_{2} $ planes in cuprate high-temperature superconductors. Despite being more piezoelectric, the dielectric tensors in these ferroelectrics are so high that the interaction decreases [but not the ratio to the also highly screened Coulomb repulsion; see Eq. (\ref{eqn:ratio_e_p})].

\begin{figure}
	\raggedleft
	\includegraphics[width=1\linewidth]{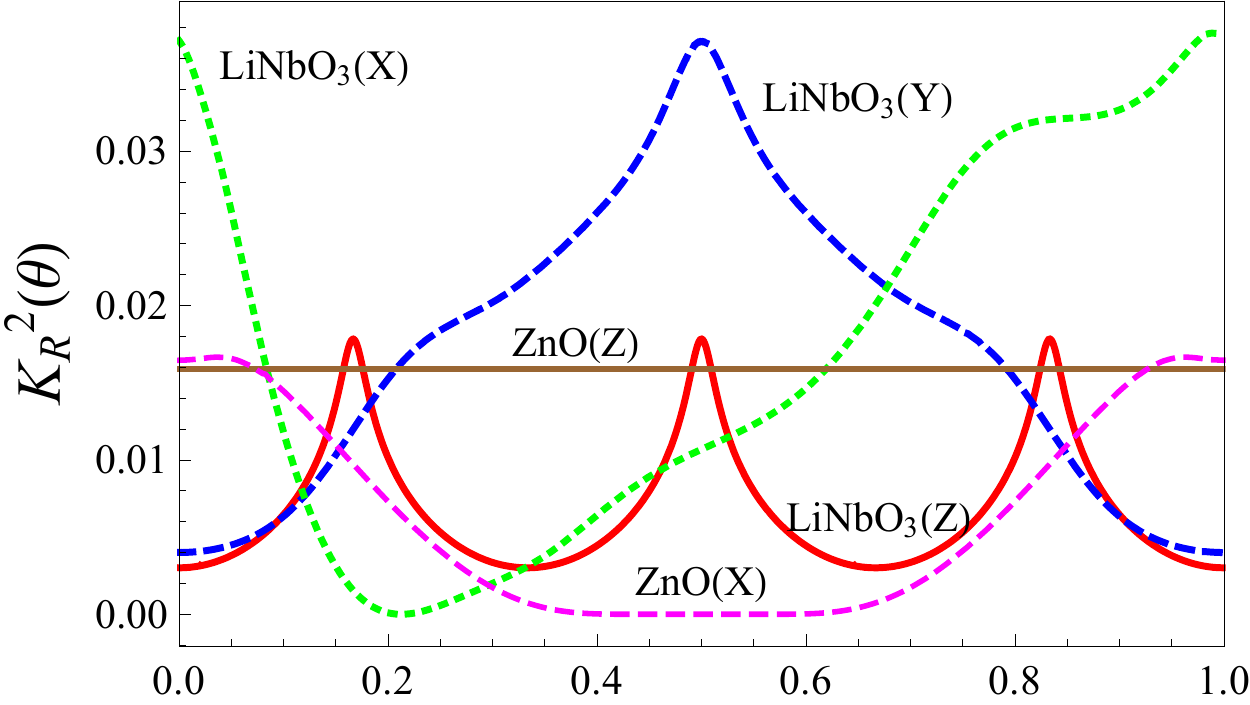}
	\vskip 0.5cm
	\includegraphics[width=0.96\linewidth]{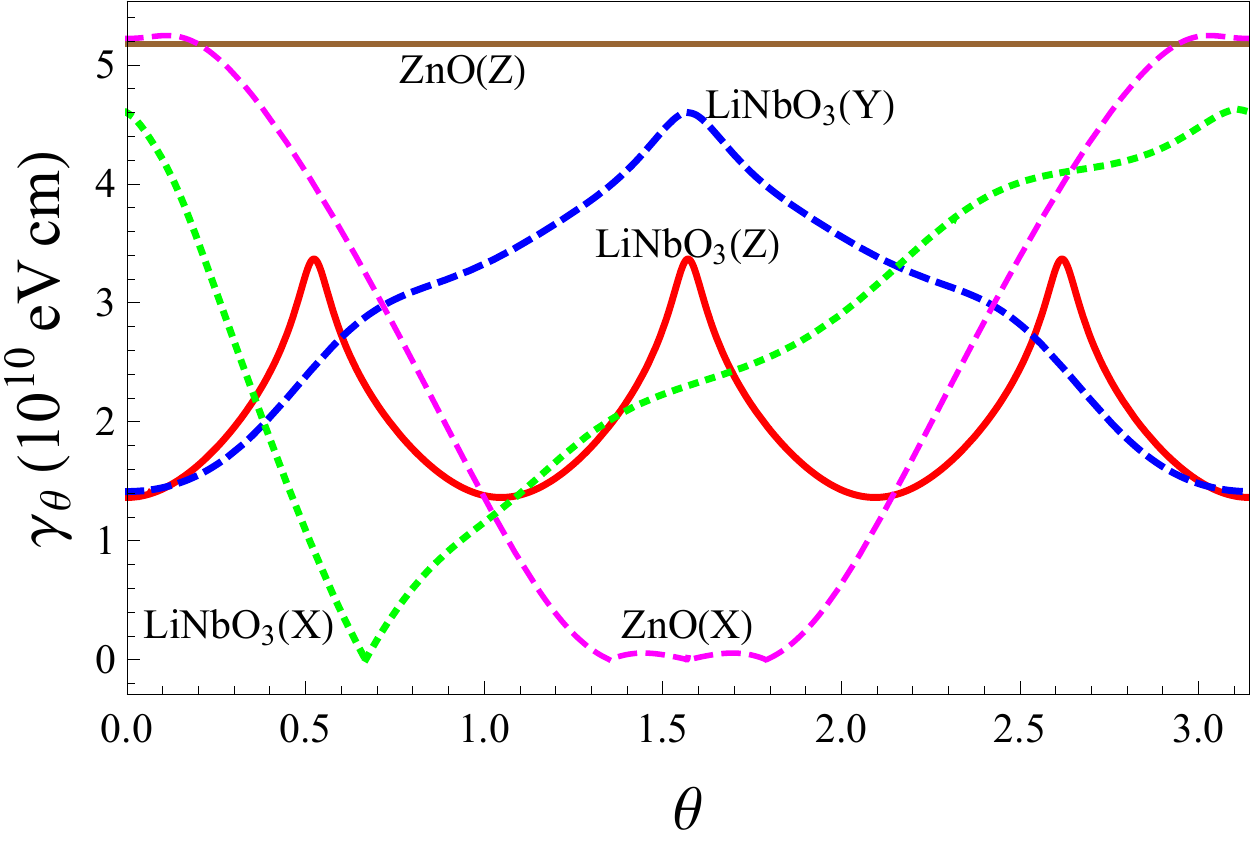}
	\caption[comentarios breves]{(Color online) Some representative magnitudes for SAW phonons and for two different materials with different symmetries ($\text{LiNbO}_3$ and ZnO), as a function of the angle within the crystal plane parallel to the cut plane, in the X-Y-Z-cuts (denoted between parentheses) chosen for all. Upper: Electromechanical coupling coefficient. Lower: Electron-phonon vertex. Data taken from Refs.~\cite{Royer2000,Auld1990,Knabchen1996} and references therein.}
	\label{fig:vertexPlot}
\end{figure}

The point group of ZnO and AlN gives isotropic couplings with the Z-cut and therefore isotropic sound velocities. On the other hand, their X and Y cuts are equivalent. This does not happen, for example, in $\text{LiNbO}_3$, whose $K_R^2(\theta)$ and vertex values in the X,Y and Z-cuts are shown in Fig. \ref{fig:vertexPlot} as an example. For some graphs of the velocities in different cuts, see for example Ref. \cite{Campbell1968}.

\section{Phonon self-energy}
\label{sec:phonon_SE}
As the piezoelectric coupling Eq.~(\ref{eqn:PAHamiltonian}) enables the transfer of energy between
carriers in graphene and the phonon modes of the substrate material, the latter
acquire an extra decay rate due to Landau damping. In order to assess the
magnitude of this effect, we proceed to estimate the self-energy of the
substrate phonons due to their interactions with the graphene carriers. Substituting the bare propagator (\ref{eqn:bare-propagator})
into  (\ref{eqn:dressedPhononPropagator}), we obtain
\begin{equation}
\tilde{G}^{\text{PA}}(\mathbf{q}, \omega) = \frac{2 \omega_{\mathbf{q}}/\hbar}{\omega^{2}-\omega_{\mathbf{q}}^{2} - 2\hbar^{-1}
\omega_{\mathbf{q}} |\gamma_{\mathbf{q}}^{\text{PA}}|^{2}\frac{\Pi_{0}(q,\omega)}{\varepsilon_{\text{RPA}}(\mathbf{q},\omega)}}~.
\end{equation}

In the phonon frequency range $\omega \sim v_s q \ll v_{F} q$ in which we will be mostly interested, we can approximate (see e.g. Ref. \cite{Li2013})
\begin{equation}
\Pi_{0}(q,\omega) \simeq - D(E_{F}) \left(1+i \frac{\omega}{v_{F}q}\right)
\end{equation}
in the RPA electron-electron dielectric function Eq.
(\ref{eqn:dielectricFunctionRPA}), so that, in terms of the parameter
$\lambda_{\rm e-ph}(\theta)$, the poles of $\tilde{G}^{\text{PA}}$ are shifted to
\begin{align}
\tilde{\omega}_{\mathbf{q}}
	&=
	\pm v_{s}q
\left(
1- \lambda_{\rm e-ph}\frac{k_{F}}{q+k_{\text{TF}}}
\right)^{\frac{1}{2}} \nonumber \\
	& \mp i\; \lambda_{\rm e-ph} \frac{v_{s}}{v_{F}} \frac{v_s k_{F}}{2} \left(\frac{q}{q+k_{\text{TF}}}\right)^2~.
\end{align}
In the long wavelength limit ($q\ll k_F$), the leading order of the ratio of the imaginary and real parts of the dressed phononic energy goes like
\begin{equation}
\left|\frac{\text{Im}(\tilde{\omega}_{\mathbf{q}})}{\text{Re}(\tilde{\omega}_{\mathbf{q}})} \right|
%	= \frac{\frac{\lambda_{\rm e-ph} \overline{\varepsilon}_{0}}{4 \alpha_{\text{fs}}}\; \frac{v_{s}}{v_{F}}}{\sqrt{1- \frac{\lambda_{\rm e-ph} \overline{\varepsilon}_{0}}{4\alpha_{\text{fs}}}}}
	\simeq \frac{1}{2} K_R^2\frac{v_s}{v_F} \left(\frac{q}{k_\text{TF}}\right)^{\pm 1}~,
	\label{eqn:relativePhononWidth}
\end{equation}
where $(q/k_\text{TF})^{\pm 1} \ll 1$, the case $k_{\rm TF}\ll q \ll k_F$ being meaningful only in those materials where $k_{\rm TF}$ is substantially smaller that $k_F$. Due to the fact that $ v_{F} / v_{s}  \sim 300$ and to the $K_R^{2}(\theta)$ values shown in Table \ref{tab:materials} for typical materials, the lifetime of the phonons can be neglected in all analyzed regimes. It can also be shown that, near the quasiparticle poles, the residue $Z_q$ is close to unity (i.e., the wave function renormalization is weak):
\begin{align}
\tilde{G}^{\text{PA}}(\mathbf{q}, \omega) \simeq Z_q \frac{2\tilde{\omega}_\mathbf{q}/\hbar}{\omega^2-\tilde{\omega}_\mathbf{q}^2} %\nonumber 
\label{G-ren-Z-q}\\
Z_q \simeq 1+ \lambda_{\rm e-ph}\frac{k_{F}}{q+k_{\text{TF}}} \, . \label{residue-q}
\end{align}

Thus in the following we can assume the substrate phonons to be well-defined, stable quasiparticles, and we will approximate the renormalized phonon propagator (\ref{G-ren-Z-q}) by the bare one, Eq. (\ref{eqn:bare-propagator}).
%%%%%%%%%%%%%%%%%%%%%%%%%%%%%%%%%%%%%%%%%%%%%%%%%%%%%%%%%%%%%%%%%%%%%%%%%%%%%%%%%%%%%%%%%%%%%%%%%%%%
\section{Electron self-energy}
\label{sec:electron_SE}
%%%%%%%%%%%%%%%%%%%%%%%%%%%%%%%%%%%%%%%%%%%%%%%%%%%%%%%%%%%%%%%%%%%%%%%%%%%%%%%%%%%%%%%%%%%%%%%%%%%%
\begin{figure}
	\centering
	\includegraphics[width=0.9\linewidth]{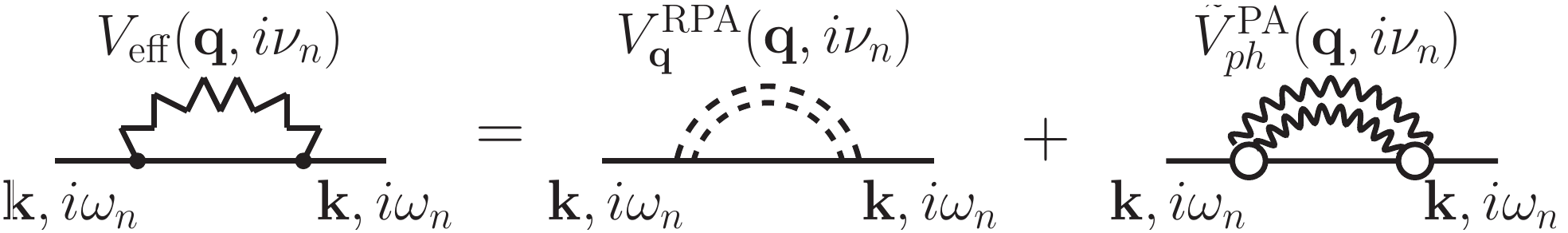}
	\caption[comentarios breves]{Electron self-energy in the $G_0W$ approximation, see Eq.~(\ref{eqn:electronSelfEnergyGW}).}
	\label{fig:ElectronSelfEnergyGW}
\end{figure}

We focus on the case of n-doped graphene ($E_F>0$) so that we will be interested in the electron self-energies at energies $\hbar \omega$ in the upper Dirac cone. With an effective electron-electron interaction $V_{\text{eff}}$ given in (\ref{eqn:V_effAlternative}), the
self-energy acquired by the charge carriers in graphene (within the $G_{0}W$
approximation, as indicated in Fig.~\ref{fig:ElectronSelfEnergyGW}) has the general form
\begin{eqnarray}
&
\Sigma_{+}(\mathbf{k},i\omega_{n})
	=
	-k_{B}T \, \sum_{s=\pm} \sum_{\mathbf{q}} \sum_{i\nu_{n}}
	F_{+s}(\mathbf{k,k+q})
\nonumber
\\
	&\times
	G_{0,s}^{\text{el}}(\mathbf{k+q},i\omega_{n}+i\nu_{n})
	V_{\text{eff}}(\mathbf{q},i\nu_{n})
,
\label{eqn:electronSelfEnergyGW}
\end{eqnarray}
where the subscript + refers to the conduction band (the calculation for $\Sigma_-$ being analogous),
the index $s=\pm$ is summed over both bands,
\begin{equation}
G_{0,s}^{\text{el}}(\mathbf{k},\omega)=(\omega - E_{\mathbf{k}s}-\mu)^{-1}
\label{eqn:bare-electron-propagator}
\end{equation}
denotes the (bare) electron propagator, $i\nu_{n}$ and $i\omega_{n}$ are, respectively, the bosonic and fermionic Matsubara frequencies, and the spinor overlap factor
\begin{equation}
F_{+s}(\mathbf{k,k+q}) = \frac{1}{2}(1+s\cos \alpha)
\label{eq:Fskkq}
\end{equation}
arises due to the sublattice structure of graphene \cite{CastroNeto2009}, $\alpha$ being the angle formed by $\mathbf{k}$ and $\mathbf{k+q}$.

Expression (\ref{eqn:V_effAlternative}) for $V_{\text{eff}}$ allows us to
separate the self-energy $\Sigma_{+}$ into contributions due to electron-electron
and electron-phonon interactions. While the former has been
considered in Refs.~\cite{Tse2007,Li2013}, the contributions of
graphene-intrinsic optical or acoustic phonons, as well as optical substrate phonons, to the electron self-energy
have been studied in Refs.~\cite{Li2013,Hwang2013}. Thus in the
present work we focus entirely on the effect of piezoelectric acoustic substrate
phonons, as expressed in the self-energy
\begin{eqnarray}
\Sigma_{+}^{\text{PA}}(\mathbf{k}&,i\omega_{n})
	= -k_{B}T \, \sum_{s=\pm} \sum_{\mathbf{q}} \sum_{i\nu_{n}}
	F_{+s}(\mathbf{k,k+q}) \nonumber
\\
	&\times
	G_{0,+}^{\text{el}}\left(\mathbf{k+q},i\omega_{n}+i\nu_{n}\right) \tilde{V}_{\text{ph}}^{\text{PA}}(\mathbf{q},i\nu_{n})
\label{sigma-plus-PA}
,
\end{eqnarray}
where
\begin{equation}
\tilde{V}_{\text{ph}}^{\text{PA}}(\mathbf{q}, \omega)
	\equiv
	\left|\frac{\gamma_{\mathbf{q}}^{\text{PA}}}{\varepsilon_{\text{RPA}}(\mathbf{q},\omega)} \right|^{2} \tilde{G}^{\text{PA}}(\mathbf{q},\omega)
\label{eqn:V_effPAPhononContribution}
.
\end{equation}

In order to sum over over Matsubara frequencies, we follow Ref. \cite{Mahan2013} and 
approximate the vertex renormalization by its static limit [see Eq. (\ref{eqn:dielectricFunctionRPAStaticLimit})] while neglecting the phonon self-energy, i.e., in Eq. (\ref{sigma-plus-PA}) we replace $\tilde{V}_{\text{ph}}^{\text{PA}}(\mathbf{q}, \omega)$ by
\begin{equation}
\bar{V}_{\text{ph}}^{\text{PA}}(\mathbf{q}, \omega)
\equiv
\left|\frac{\gamma_{\mathbf{q}}^{\text{PA}}}{\varepsilon_{\text{RPA}}(\mathbf{q}, 0)} \right|^{2} G_0^{\text{PA}}(\mathbf{q},\omega) \, .
\end{equation}
We arrive at the following retarded self-energy:
\begin{eqnarray}
&
\Sigma_{+}^{\text{PA}}(\mathbf{k},\omega)
	=
%	\lim\limits_{ \eta \rightarrow 0^{+}}
	{\huge \sum_{s=\pm} \int } \frac{d\mathbf{q}}{(2\pi)^{2}} \left|\frac{\gamma_{\mathbf{q}}^{\text{PA}}}{\varepsilon_{\text{RPA}}(\mathbf{q},0)} \right|^{2}
F_{+s}(\mathbf{k,k'})
\nonumber
\\
& \times \left[\frac{n_{B}(\hbar \omega_{\mathbf{q}}) + n_{F}(\epsilon_{k's})}{\hbar\omega + \hbar \omega_{\mathbf{q}}-\epsilon_{k's} +i0^+} + \frac{n_{B}(\hbar \omega_{\mathbf{q}})+1 - n_{F}(\epsilon_{k's})}{\hbar\omega - \hbar \omega_{\mathbf{q}}-\epsilon_{k's} +i0^+}\right]
\label{eqn:electronSelfEnergyRetarded}
,
\end{eqnarray}
where $\mathbf{k'}$ stands for $\mathbf{k' \equiv k+q}$,
\begin{eqnarray}
n_{B}(\hbar \omega_{\mathbf{q}}) &=& \left[\exp\left(\frac{\hbar \omega_\mathbf{q}}{k_{B}T}\right)-1\right]^{-1} \, , \\ n_{F}(\epsilon_{k's})&=& \left[\exp\left(\frac{\epsilon_{k's}}{k_{B}T}\right)
+1\right]^{-1}
\end{eqnarray}
denote the Bose and Fermi distributions, respectively, and the energies
$\epsilon_{ks} = E_{ks}-\mu$ are taken relative to the chemical potential.
We proceed by evaluating the real and imaginary parts of Eq.~(\ref{eqn:electronSelfEnergyRetarded}) separately. Hereafter,
we assume $T \ll T_F$ so that the zero-temperature RPA dielectric function can be used \cite{Wunsch2006}. Since $\mu \simeq E_F$, we can write
\begin{equation}
\epsilon_{ks}=\hbar v_F(ks-k_F)~.
\end{equation}
%, with $\lambda \equiv \lambda_{\text{e-ph}}$.
%%%%%%%%%%%%%%%%%%%%%%%%%%%%%%%%%%%%%%%%%%%%%%%%%%%%%%%%%%%%%%%%%%%%%%%%%%%%%%%%%%%%%%%%%%%%%%%%%%%%
\subsection{Imaginary part}
\label{subsec:electron_SE_im}
%%%%%%%%%%%%%%%%%%%%%%%%%%%%%%%%%%%%%%%%%%%%%%%%%%%%%%%%%%%%%%%%%%%%%%%%%%%%%%%%%%%%%%%%%%%%%%%%%%%%
The imaginary part of Eq. (\ref{eqn:electronSelfEnergyRetarded}) acquires the form
\begin{eqnarray}
& \text{Im}\, \Sigma_{+}^{\text{PA}}(\mathbf{k},\omega) = - \pi {\huge \sum\limits_{s=\pm}^{} \sum\limits_{t=\pm} \int } \frac{d\mathbf{q}}{(2\pi)^{2}} \left|\frac{\gamma_{\mathbf{q}}^{\text{PA}}}{\varepsilon_{\text{RPA}}(\mathbf{q},0)} \right|^{2} \frac{1+s\cos \alpha}{2} \nonumber \\
& \times \left[n_{F}(\hbar \omega_{\mathbf{q}} + t \hbar
\omega) + n_{B}(\hbar \omega_{\mathbf{q}})\right] \delta (\hbar\omega + t\hbar \omega_{\mathbf{q}} - \epsilon_{k's}),~~~~ \label{eqn:electronSelfEnergyImPart}
\end{eqnarray}
where $t$=$\pm 1$ corresponds to the absorption or emission of a phonon,
respectively.
%As before, $\alpha$ denotes the angle between $\mathbf{k}$ and
%$\mathbf{k+q}$.

Setting $\omega = \epsilon_{k+}$ in Eq.~(\ref{eqn:electronSelfEnergyImPart}), that is, considering the on-shell self-energy,
yields the value $\hbar/(2 \tau)$ for the decay width of charge-carriers with wavevector $\mathbf{k}$. Here we are assuming that the renormalization  of the Fermi energy $\Delta E_F = \text{Re} \Sigma^\text{PA}(k_F,0)$, as given by the pole of the dressed electron propagator,  is tiny, as can be checked in the next section [see Eq. (\ref{eqn:electronSelfEnergyRePartComputed}) and related ones]. To obtain analytical expressions for the asymptotic behaviors of the on-shell self-energy,
we introduce the quasi-elastic approximation
\begin{equation}
\delta(\epsilon_{k+} +t\hbar \omega_{\mathbf{q}} - \epsilon_{k's}) \simeq \delta(\epsilon_{k+}  - \epsilon_{k's})
\end{equation}
in Eq.~(\ref{eqn:electronSelfEnergyImPart}),
which is well justified since $v_{F}/v_{s} \sim 300$. 
As we are working with $k_F>0$, the $s=-$ term is null. Hereafter, $\epsilon_{k}$ will be equivalent to $\epsilon_{k+}$, so that 
\begin{equation}
\epsilon_{k}=\hbar v_F(k-k_F) \, .
\end{equation}
For magnitude estimates we will assume $\epsilon_k>0$.

The relevant scale for finite temperature effects in graphene, where carrier
densities are much smaller than in conventional metals,
%these low-doping systems
is the Bloch-Gr\"uneisen temperature $T_{\text{BG}}$, defined as the scale of
the acoustic phonons in the Fermi sea,
%(not the Debye zone as in the
%usual metals with $Z$ electrons per atom), where $k_{B}T_{\text{BG}} = 2\hbar v_{s} k_{F}$.
\begin{align}
k_{B}T_{\text{BG}} \equiv 2\hbar v_{s} k_{F} \,
.
\label{eqn:TBG_def}
\end{align}
Then at zero temperature (by which we mean $T\ll \epsilon_k/k_B, T_{\rm BG}$), $n_{F}$ in (\ref{eqn:electronSelfEnergyImPart}) becomes a step function which cuts off the
momentum integration, while $n_{B}$ vanishes.
Then, in the limit
$\epsilon_{k} \ll \hbar v_s k_{\rm TF}$ (for which the largest contributing $q$ in (\ref{eqn:electronSelfEnergyImPart}) is 
%$q \sim v_F|k-k_F|/v_s \ll k_{\rm TF}$
$q \sim \epsilon_{k}/\hbar v_s$ so that we can assert $q \ll k_{\rm TF}$)
the quasiparticle lifetime decays as a $\epsilon_{k}^3$ near the Fermi surface while depending on the direction of the $\mathbf{k}$ vector:
\begin{align} &-\text{Im}\,
\Sigma_{+}^{\text{PA}}(\mathbf{k},\epsilon_{k}) \simeq
\frac{1}{6\pi}\frac{|\gamma_{\perp}^{\text{PA}}|^{2}}{\hbar
v_{F} k_{\text{TF}\perp}^{2}}\frac{\epsilon_{k}^{3}}{(\hbar
v_{s\perp})^{3}} \nonumber \\ &= \frac{\lambda_{\perp}}{24}
\left(\frac{k_F}{k_{\rm TF\perp}}\right)^{2}\left(\frac{v_{F}}{v_{s\perp}}\right)^{2}
\left(\frac{\epsilon_{k}}{E_{F}}\right)^{3} \hbar v_{F}
k_{F}~, \label{eqn:electronSelfEnergyImPartZeroTemperature}
\end{align}
where all the substrate related constants, like $\lambda_{\perp} \equiv \lambda_\text{e-ph}(\theta_{\perp \mathbf{k}})$ of Eq.~(\ref{eqn:lambdaDefinition}), have
to be taken in the direction $\theta_{\perp \mathbf{k}}$ perpendicular to $\mathbf{k}$.
Hereafter $\lambda$ will replace $\lambda_{\perp}$ whenever substrate related constants are assumed to be direction independent.
The fast $\epsilon_{k}^3$ decrease (as $\epsilon_{k} \rightarrow 0$) is due to the vertex renormalization, since  $\varepsilon_{\text{RPA}}$ in
Eq.~(\ref{eqn:electronSelfEnergyImPart}) diverges for
%$v_{F}(k-k_{F}) \leq v_{s} k_{\text{TF}}$
$q\ll k_{\text{TF}}$
[see Eq. (\ref{eqn:dielectricFunctionRPAStaticLimit})].

For $\epsilon_{k} \gg \hbar v_s k_{\text{TF}}$ we obtain the result
\begin{align} 
-\text{Im}\, \Sigma_{+}^{\text{PA}}(k,\epsilon_{k}) & \simeq \frac{\lambda}{4} \hbar v_s k \int\limits_{0}^{1} \frac{y^2\sqrt{1-y^{2}}}{\left(y+\frac{k_\text{TF}}{2k}\right)^{2}} \, dy  \nonumber \\
& = \frac{\lambda \hbar v_s k}{4}  f\left(\frac{k_\text{TF}}{2k}\right)\, ,
\label{eqn:T0-energy-larger-vskTF} \\
f(x) =3x & +\frac{\pi}{4}(1-6x^2) + \frac{(3x^3-2x)\, \text{acosh}(x^{-1})}{\sqrt{1-x^2}} \label{eqn:functionFF} \, .
\end{align}
This admits two regimes: for $\hbar v_s k_{\text{TF}} \ll \epsilon_k \ll E_F$,
\begin{equation}
	-\text{Im}\, \Sigma_{+}^{\text{PA}}(k,\epsilon_{k}) \simeq \frac{\lambda k_B T_{\rm BG}}{8}  f\left(\frac{2\alpha_{\text{fs}}}{\overline{\varepsilon}_{0}}\right) \, ,
\label{eqn:hvskTF-eps-EF}
\end{equation}
while for $\epsilon_k \gg E_F$ we obtain
\begin{equation}
-\text{Im}\, \Sigma_{+}^{\text{PA}}(k,\epsilon_{k}) \simeq 
\frac{\pi \lambda v_s}{16 v_F} \epsilon_k \, .
\label{eqn:hvskTF-EF-eps}
\end{equation}

Returning to the low energy ($\epsilon_{k} \ll \hbar v_s k_{\text{TF}}$) regime [see Eq. (\ref{eqn:electronSelfEnergyImPartZeroTemperature})], we note that, without the vertex screening effect [that is, setting  $\varepsilon_{\text{RPA}}\to 1$
in Eq.~(\ref{eqn:electronSelfEnergyImPart})], instead of the $\epsilon_k^3$ behavior one would find
the linear $\epsilon_{k}$ dependence characteristic of a marginal Fermi liquid,
\begin{equation}
- \text{Im}\, \Sigma_{+\text{(no scr)}}^{\text{PA}}(\mathbf{k},\epsilon_{k}) \simeq \frac{\lambda_{\perp}}{8} \epsilon_{k}
% = \frac{\lambda_{\perp}}{8} \frac{\epsilon_{k}}{E_{F}}\;\hbar v_{F} k_{F}
\label{eqn:electronSelfEnergyImPartZeroTemperatureNoScreening} \,,
\end{equation}
which (for materials such that $k_{\rm TF} \ll k_{F}$) behaves similarly to the true self-energy in the range 
$\hbar v_s k_{\rm TF} \ll \epsilon_{k} \ll \hbar v_s k_{F}$, since $\varepsilon_{\text{RPA}}$ tends to unity for the momenta $q \gg k_{\rm TF}$ dominating the integral
(\ref{eqn:electronSelfEnergyImPart}). We will see however that that a small offset remains due to the contribution of the screened low-$q$ values ($q \ll k_{\rm TF}$).
%\todo{[[The previous sentences and the previous limit should be checked. What is exactly its regime of validity? Does it hold for materials such that $k_TF < k_F$ (case of ZnO)? If so, what is its regime of validity? This is important for understanding Fig. 5a discussed in sec. V]]}
Here and in the following, we remove the subindex $\perp$ from the anisotropic parameters in those expressions where only their order of magnitude matters.

Table \ref{tab:materials} shows representative angle-independent material parameters, including those that will be used for the numerical calculations discussed in Section \ref{sec:numerical}.
From Eqs. (\ref{eqn:electronSelfEnergyImPartZeroTemperature}),(\ref{eqn:electronSelfEnergyImPartZeroTemperatureNoScreening}) and the parameter values shown in Table \ref{tab:materials}, it is safe to conclude that, at zero temperature, the damping rate due to electron-phonon coupling is always much smaller than $\epsilon_{k}$. Thus the single-electron quasiparticles near the Fermi surface are well defined.

So far we have assumed zero temperature, i.e., $k_B T \ll \epsilon_{k}$.
At nonzero temperatures, the vertex renormalization is fundamental to avoid
logarithmic divergences. These occur for the unscreened self-energy at any nonzero temperature due to the divergent contribution of small $q$ values.
Focusing on the correctly screened self-energy, we consider first the nonzero, low-temperature limit
$\epsilon_{k} \ll k_B T \ll 2 \hbar v_s k_{\rm TF} 
%\lesssim
, k_{B}T_{\text{BG}}$. Again, only the perpendicular-to-$\mathbf{k}$ substrate-related constants appear. We obtain
\begin{align}
-\text{Im}\, \Sigma_{+}^{\text{PA}}(k,\epsilon_{k}) \simeq \lambda_{\perp} k_{B} T \left(\frac{k_F}{k_{\rm TF\perp}}\right)^{2}\left(\frac{T}{T_{\text{BG}\perp}}\right)^{2} \frac{7\zeta(3)}{2}~, \label{eqn:electronSelfEnergyImPartLowTemperature}
\end{align}
with $7\zeta(3)/2\simeq 4.21$. The essential independence from $k$ of the lifetime (which allows for the replacement $k \simeq k_F$) is a general property of the case $\epsilon_k \ll T$.
In those materials where $\varepsilon_0$ is so high that $k_\text{TF} \ll k_F$ and therefore a temperature regime exists such that $\epsilon_{k}\ll \hbar v_s k_\text{TF} \ll k_B T \ll k_B T_\text{BG}$, 
the $T^3$ law is replaced by a $\sim T \log T$ behavior. Specifically, the asymptotic expression reads
%\todo{[[Ivar says he can compute the precise prefactor of T log T]]}
\begin{align}
-\text{Im}\, \Sigma_{+}^{\text{PA}}(k,\epsilon_{k})\simeq \lambda_{\perp} k_{B} T \left(\frac{k_F}{k_{\rm TF\perp}}\right)^{2}\log \left(\frac{\hbar v_s k_\text{TF}}{k_B T}\right)~. 
\label{eqn:ImSTlogT}
\end{align}
%In that case, the higher $\varepsilon_0$ is, the more $\lambda_{\perp}$ decreases, so that the effect ends up being small. 

The high-temperature limit ($T_{\text{BG}} \ll T$, while only $\epsilon_{k}\ll E_F$ is required),
where phonons are nondegenerate, yields
\begin{align}
-\text{Im}\, \Sigma_{+}^{\text{PA}}(k,\epsilon_{k}) & \simeq \frac{\lambda}{4} k_{B} T \int\limits_{0}^{1} \frac{y\sqrt{1-y^{2}}}{\left(y+\frac{k_\text{TF}}{2k}\right)^{2}} \, dy  \nonumber \\
& = \frac{\lambda k_{B} T}{4}  g\left(\frac{k_\text{TF}}{2k}\right)\, ,
\label{eqn:electronSelfEnergyImPartHighTemperature} \\
g(x) = -2 & +\pi x + \frac{(1-2x^2)\, \text{acosh}(x^{-1})}{\sqrt{1-x^2}} \label{eqn:functionF} \, .
\end{align}

The logarithmic divergence of the function $g$ at $x \rightarrow 0$ becomes relevant in the limit $k \gg k_{\text{TF}}$, where
\begin{equation}
-\text{Im}\, \Sigma_{+}^{\text{PA}}(k,\epsilon_{k})  \simeq
\frac{\lambda k_{B}T}{4}\left[\log \left(\frac{4k}{k_{\text{TF}}}\right)-2 \right]~.
\label{eqn:lim-k-large}
\end{equation}

Comparing Eqs.~(\ref{eqn:electronSelfEnergyImPartZeroTemperature}),
(\ref{eqn:electronSelfEnergyImPartLowTemperature}), and
(\ref{eqn:electronSelfEnergyImPartHighTemperature}) with the corresponding limiting expressions
for the electron self-energy induced by the graphene-intrinsic
deformation-potential acoustic (DA) phonons \cite{Li2013}, we see below that, for an important range of parameter values, the inverse lifetime
is dominated by the piezoelectric substrate phonons.

For our estimates we borrow
$\Sigma_{+}^{\text{DA}}(k,\epsilon_{k})$ from Ref. \cite{Li2013}.
Specifically, with a deformation constant $D\simeq 25$ eV, and taking
$k_{F} = [k_F]\,10^{6}\, \text{cm}^{-1}$ (this momentum unit corresponds to a density of $k_F^2/\pi\simeq 3.2 \times 10^{11}$ cm$^{-2}$), one obtains from (\ref{eqn:electronSelfEnergyImPartZeroTemperature})
\begin{equation}
\frac{\text{Im}\, \Sigma_{+}^{\text{PA}}(k,\epsilon_{k})}{\text{Im}\, \Sigma_{+}^{\text{DA}}(k,\epsilon_{k})} \simeq \frac{20}{[k_F]^{2}} \lambda \overline{\varepsilon}_{0}^{2} \frac{\epsilon_{k}}{1\; \text{meV}} ~,
\end{equation}
for $k_BT \ll \epsilon_{k} \ll \hbar v_s k_{\rm TF}$. 
%This ratio is bigger than 1 for usual piezoelectric materials and not very low $\epsilon_{k}$ (less than $\sim 0.1 \, \text{meV}$).

Likewise, at nonzero temperatures ($\epsilon_{k}/k_B \ll T \ll T_{\text{BG}}$), we have from (\ref{eqn:electronSelfEnergyImPartLowTemperature})
\begin{equation}
\frac{\text{Im}\,
\Sigma_{+}^{\text{PA}}(k,\epsilon_{k})}{\text{Im}\, \Sigma_{+}^{\text{DA}}(k,\epsilon_{k})} \simeq \frac{100}{[k_F]^{2}} \lambda \overline{\varepsilon}_{0}^{2} \frac{k_{B}T}{1\; \text{meV}}~.
\label{eqn:ratioElectronSelfEnergyImPartPADA-lowT}
\end{equation}

Finally, at high temperatures ($\epsilon_{k}/k_B \ll T_{\text{BG}}\ll T$), one obtains from (\ref{eqn:electronSelfEnergyImPartHighTemperature}) the $k$-independent ratio
\begin{equation}
\frac{\text{Im}\,
\Sigma_{+}^{\text{PA}}(k,\epsilon_{k})}{\text{Im}\, \Sigma_{+}^{\text{DA}}(k,\epsilon_{k})} \simeq \frac{35 \, g\left(\frac{2\alpha_{\text{fs}}}{\overline{\varepsilon}_{0}}\right)}{[k_F]} \lambda ~.  
\label{eqn:ratioElectronSelfEnergyImPartPADA}
\end{equation}
%We recall here that the function $f(x)$ diverges at $x\rightarrow 0$ [see Eq. %(\ref{eqn:lim-k-large})].

From these ratios we conclude that piezoelectric acoustic phonons can dominate over deformation acoustic phonons in an appreciable range of realistic material parameters, especially for small carrier concentrations. 
%($[k_F] \gtrsim 10$).
The smaller value of $D \,\simeq 6.8$ eV also found in the literature \cite{Kaasbjerg2012,Zhang2013} would further increase the relative importance of piezoelectric phonons against intrinsic ones.

\subsection{Real part}
\label{subsec:electron_SE_re}
%%%%%%%%%%%%%%%%%%%%%%%%%%%%%%%%%%%%%%%%%%%%%%%%%%%%%%%%%%%%%%%%%%%%%%%%%%%%%%%%%%%%%%%%%%%%%%%%%%%%
%
For the real part of the self-energy we have, from Eq. (\ref{eqn:electronSelfEnergyRetarded}),
\begin{eqnarray}
\text{Re}\, &\Sigma_{+}^{\text{PA}}(\mathbf{k},\omega)  =
\sum_{s=\pm}
\int  \frac{d\mathbf{q}}{(2\pi)^{2}}
\left|\frac{\gamma_{\mathbf{q}}^{\text{PA}}}{\varepsilon_{\text{RPA}}(\mathbf{q},0)}
\right|^{2} F_{+s}(\mathbf{k,k'})
\nonumber
\\
& \times\left[\frac{n_{B}(\hbar \omega_{\mathbf{q}}) + n_{F}(\epsilon_{k's})}{\hbar\omega + \hbar\omega_{\mathbf{q}}-\epsilon_{k's}} + \frac{n_{B}(\hbar \omega_{\mathbf{q}})+1 - n_{F}(\epsilon_{k's})}{\hbar\omega - \hbar \omega_{\mathbf{q}}-\epsilon_{k's}}\right]~,
\label{eqn:electronSelfEnergyRePart}
\end{eqnarray}
where the denominators are to be understood as principal values. Unlike for many-body effects directly caused by the electron-electron interaction, this phonon contribution to the electron self-energy tends to be
negligibly small compared to the Fermi energy. However, its derivatives are large. As a result, the phonon-induced contributions to the Fermi velocity renormalization are larger than those stemming from the direct electron-electron interactions.

Since $\partial
\text{Re}\,\Sigma_{+}^{\text{PA}}(\mathbf{k},\omega) /\partial (v_F\mathbf{k})$ is, by a factor of $v_{s}/v_{F}$,
smaller than
$\partial \text{Re}\,\Sigma_{+}^{\text{PA}}(\mathbf{k},\omega) /\partial \omega$ (see Ref. \cite{Migdal1958}), it suffices to focus on the frequency derivative, in contrast to
the case of electron-electron interactions, where both derivatives matter \cite{Mahan2013,DasSarma2007}.
We thus approximate
\begin{equation}
\tilde{v}_{F}(\hat{\mathbf{k}}) = v_{F}\left[1-
\frac{\partial \text{Re}\, \Sigma_{+}^{\text{PA}}(\hat{\mathbf{k}}k_{F},\omega)}
{\partial \omega}\bigg|_{\omega=0}\right]^{-1} \\
%\frac{\tilde{v_{F}}(k)}{v_{F}}=\frac{1}{1- \text{Re}
%\frac{\partial \Sigma_{+}^{\text{(a)}}(\mathbf{k},\omega)}{\partial
%\omega}_{|\omega=\epsilon_{k}}} ~,
\label{eqn:vF-renor-anisotropic}
\end{equation}
for the (direction dependent)
renormalization of the Fermi velocity in graphene induced by piezoelectric
acoustic substrate phonons.

For further analysis, it us useful to separate Eq.~(\ref{eqn:electronSelfEnergyRetarded})
into three terms, 
%\cite{Mahan2013}
%\begin{equation}
%\text{Re}\,\Sigma_{+}^{\text{PA}} = \text{Re}\,\Sigma_{+}^{\text{(ph)}} + %\text{Re}\,\Sigma_{+}^{\text{(el)}} + \text{Re}\,\Sigma_{+}^{\text{(vac)}} \, ,
%\label{eqn:electronSelfEnergyRePart_Ph_a_b}
%\end{equation}
\begin{equation}
\Sigma_{+}^{\text{PA}} = \Sigma_{+}^{\text{(ph)}} + \Sigma_{+}^{\text{(el)}} + \Sigma_{+}^{\text{(vac)}} \, ,
\label{eqn:electronSelfEnergyRePart_Ph_a_b}
\end{equation}
where $\Sigma_{+}^{\text{(ph)}}$ contains just the Bose factor $n_{B}(\hbar \omega_{\mathbf{q}})$,
$\Sigma_{+}^{\text{(el)}}$ the Fermi factor $n_{F}(\epsilon_{k's})$, and $\Sigma_{+}^{\text{(vac)}}$ the remaining vacuum term. As in the previous subsection, in the following angle-independent material parameters are assumed.

The real part of $\Sigma_{+}^{\text{(vac)}}$ at $\omega=0$ is independent of the Fermi energy:
\begin{align}
\text{Re}\, \Sigma_{+}^{\text{(vac)}}(k,0) \simeq -
\frac{\lambda}{16} \frac{v_{s}}{v_{F}} \left[\hbar v_{F} k_{c}
+ \hbar v_{F} k \log \left(\frac{k_{c}-k}{k}\right)\right]\, , \label{eqn:electronSelfEnergyRePartComputed}
\end{align}
where $k_{c}$ is a cutoff momentum of the order of the inverse lattice spacing. Because of the small prefactor,
$\text{Re}\,\Sigma_{+}^{\text{(vac)}}(k_F,0)$ represents a weak correction to the chemical potential for all relevant carrier densities, even for $k_c \gg k_F$. We will see that its derivative can also be neglected because $\partial_\omega \text{Re}\,\Sigma_{+}^{\text{(vac)}}(k_F,0) \simeq (\lambda/4) (v_{s}/v_{F}) (1+\log k_F/k_c)   \ll \partial_\omega \text{Re}\,\Sigma_{+}^{\text{(el)}}(k_F,0)$.

At temperatures $T\ll T_{\text{BG}}$ the term containing the Bose factors $\text{Re}\,\Sigma_{+}^{\text{(ph)}}$ is exponentially small, while at temperatures $T\gg T_{\text{BG}}$ it does not grow larger than a factor $T/T_{\text{BG}}$ times the expression in Eq. (\ref{eqn:electronSelfEnergyRePartComputed}). Hence we can also neglect $\partial_\omega \text{Re}\,\Sigma_{+}^{\text{(ph)}}$.

Thus the only term that can affect the electronic properties is
$\text{Re}\,\Sigma_{+}^{\text{(el)}}(\mathbf{k},\omega)$, which is likewise small in magnitude, at most twice the term shown in Eq. (\ref{eqn:electronSelfEnergyRePartComputed}), but has a large
derivative. Note that here the quasi-elastic approximation ($\hbar \omega_{\mathbf{q}} \ll \epsilon_{k'}$) is not informative, since $\text{Re}\,\Sigma_{+}^{\text{(el)}}(\mathbf{k},\omega)$ vanishes when $\hbar \omega_{\mathbf{q}}$ is set to zero.

The integral
\begin{eqnarray}
& \frac{\partial \text{Re}\,\Sigma_{+}^{\text{(el)}}(\hat{\mathbf{k}} k_{F},\omega)}{\partial \omega}\bigg|_{\omega=0}=
-{\huge \sum_{s=\pm} \int } \frac{d\mathbf{q}}{(2\pi)^{2}} \\
& \times\left|\frac{\gamma_{\mathbf{q}}^{\text{PA}}}{\varepsilon_{\text{RPA}}} \right|^{2} F_{+s}(\mathbf{k,k'})n_{F}(\epsilon_{k's})\left[\frac{1}{( \epsilon_{k's}-\hbar\omega_{\mathbf{q}})^{2}} - \frac{1}{(\epsilon_{k's}+\hbar \omega_{\mathbf{q}})^{2}}\right] \nonumber
\end{eqnarray}
can be computed by
changing variables ($d\mathbf{q} \rightarrow d\mathbf{k'}$, with $\mathbf{k'} = \mathbf{k+q}$) and
performing the radial integral first by parts, with 
\begin{equation}
u = k'n_{F}(\epsilon_{k's})\, , \,\,\, 
dv = \frac{dk'}{(\epsilon_{k's}\pm \hbar v_{s}q)^{2}} 
\, . \nonumber
\end{equation}
We arrive at a direction-dependent expression which integrates over the Fermi surface:
\begin{eqnarray}
\frac{\partial \text{Re}\,\Sigma_{+}^{\text{(el)}}(\hat{\mathbf{k}} k_{F},\omega)}{\partial \omega}\bigg|_{\omega=0} = & \nonumber \\
-\int\limits_{0}^{2\pi} \frac{d\alpha}{\hbar v_{F}(2\pi)^{2}}  \left|\frac{\gamma_{\mathbf{k+q}}^{\text{PA}}}{\varepsilon_{\text{RPA}}} \right|^{2} & F_{+s}(\mathbf{k,k+q})\frac{2}{\hbar \omega_{\mathbf{q}}} \, ,
\end{eqnarray}
where, as in (\ref{eq:Fskkq}), $\alpha$ is the angle between $\mathbf{k}$ and $\mathbf{k+q}$.

After further averaging over the Fermi surface ($\hat{\mathbf{k}}$ directions), the ratio (\ref{eqn:vF-renor-anisotropic}) becomes similar to the temperature prefactor of the high-temperature damping (\ref{eqn:electronSelfEnergyImPartHighTemperature}),
\begin{equation}
\tilde{v}_{F} = \frac{v_{F}}{1+\frac{\lambda}{4\pi}f\left(\frac{k_\text{TF}}{2k_F}\right)}
%=\frac{v_{F}}{1+\frac{K_R^2 r_s}{\pi}f\left(2r_s\right)}
~,
\label{eqn:angle-averaged-renormalized-vF}
\end{equation}
where, we recall, all variables are angle averaged. Inspection of Eq. (\ref{eqn:angle-averaged-renormalized-vF}) shows that the renormalization of the Fermi velocity cannot exceed 3\% even for $K_R^2\sim 1$, and $K_R$ is usually much smaller. The result shown in Eq. (\ref{eqn:angle-averaged-renormalized-vF}) permits us to confirm the validity of neglecting the vacuum and phonon self-energy parts. A more accurate estimate of the ratios between derivatives yields 
$\partial_\omega \text{Re}\,\Sigma_{+}^{\text{(vac)}} /\partial_\omega \text{Re}\,\Sigma_{+}^{\text{(el)}} ={\cal O}(v_s/v_F) \ll 1$, while $\partial_\omega \text{Re}\,\Sigma_{+}^{\text{(ph)}} /\partial_\omega \text{Re}\,\Sigma_{+}^{\text{(el)}}$ is ${\cal O}(v_s/v_F)$ for $T\ll T_{\rm BG}$ and ${\cal O}(T/T_F)$ for $T_{\rm BG} \ll T \ll T_F$.

%%%%%%%%%%%%%%%%%%%%%%%%%%%%%%%%%%%%%%%%%%%%%%%%%%%%%%%%%%%%%%%%%%%%%%%%%%%%%%%%%%%%%%%%%%%%%%%%%%%%
\subsection{Electron mobility}
\label{subsec:mobilities}

Within Boltzmann transport theory, the momentum (or transport) relaxation time
$\tau_{+\text{tr}}(\mathbf{k})$ (where the subscript denotes
\textquotedblleft transport\textquotedblright~ and + denotes the band) is calculated analogously to the inverse lifetime in Sec.~\ref{subsec:electron_SE_im},
%the inverse of the
%on-shell imaginary part of the electron self-energy
%(\ref{eqn:electronSelfEnergyImPart}),
but with an extra angular factor
$(1-\cos \alpha) 
%= 2 \sin^{2}(\alpha/2) 
= q^{2}/2k^{2}$ in the integrand,
which increases the weight of large-angle scattering processes. Specifically, Eq.(\ref{eqn:electronSelfEnergyImPart}) is replaced by
\begin{eqnarray}
& \frac{\hbar}{2\,\tau_{+\text{tr}}^{\text{PA}}(\mathbf{k})} = \pi {\huge \sum\limits_{s=\pm}^{} \sum\limits_{t=\pm} \int } \frac{d\mathbf{q}}{(2\pi)^{2}} \frac{q^{2}}{2k^{2}}\left|\frac{\gamma_{\mathbf{q}}^{\text{PA}}}{\varepsilon_{\text{RPA}}(\mathbf{q},0)} \right|^{2} \frac{1+s\cos \alpha}{2} \nonumber \\
& \times \left[n_{F}(\hbar \omega_{\mathbf{q}} + t \epsilon_{k}) + n_{B}(\hbar \omega_{\mathbf{q}})\right] \delta (\epsilon_{k+} - \epsilon_{k's})~, \label{eqn:momentumRelaxationTime}
\end{eqnarray}
where the quasielastic approximation has been made.
The inclusion of this additional $q^2$ factor in the integrand improves the quasielastic approximation, changes the power law scaling
at low temperatures (by generating an extra factor $T^2$), and corrects the lifetime with a constant factor at temperatures greater than $T_{\rm BG}$.
%temperatures (in the phonon Bloch-Gr\"uneisen scale):

For quasiparticle energies such that  ($\epsilon_{k}/k_B \ll T$),
we find (after angle averaging) results that are essentially independent of $\epsilon_{k}$, i.e., $\tau_{+\text{tr}}^{\text{PA}}(k)\simeq \tau_{+\text{tr}}^{\text{PA}}(k_{F})$. In the low, yet nonzero temperature regime $\epsilon_{k}/k_B \ll T \ll 2 r_sT_{\text{BG}}$, we obtain
\begin{eqnarray}
& \frac{\hbar}{2\,\tau_{+\text{tr}}^{\text{PA}}(k_{F})} \simeq \frac{\lambda k_{B} T}{8}  \left(\frac{k_{B}T}{\hbar v_{s} k_{\text{TF}}}\right)^{4}\frac{k_{\text{TF}}^{2}}{k_{F}^{2}} \int\limits_{0}^{\infty} dx \; x^{4} \text{csch}(x) \nonumber \\
& = \frac{\lambda k_{B} T}{8} \left(\frac{\overline{\varepsilon}_{0}}{\alpha_{\text{fs}}}\right)^{2}\left(\frac{T}{T_{\text{BG}}}\right)^{4} \frac{93\zeta(5)}{2} \label{eqn:tau-trans-T5-law}
\end{eqnarray}
($93\zeta(5)/2\simeq 48.2$) which should be compared to Eq.~(\ref{eqn:electronSelfEnergyImPartLowTemperature}). 
The shift from a $T^3$ to a $T^5$ behavior is due to the transport-induced reduced weight (by a factor $q^2/2k_F^2$) of the low $q$ values dominating the inverse transport lifetime at low temperatures. 

If the vertex screening is neglected, we still obtain a convergent result, despite the temperature being nonzero,
because the low-$q$ divergence is already suppressed by the transport-associated angular weighting factor. We obtain
\begin{equation}
\frac{\hbar}{2\,\tau_{+\text{tr}}^{\text{PA}}(k_{F})_\text{(no scr)}} \simeq \frac{\lambda 7 \zeta(3)}{4} k_{B} T \left(\frac{T}{T_{\text{BG}}}\right)^{2}~, 
\label{eqn:momentumRelaxationTimeNoScreening}
\end{equation}
and recall that the non-transport equivalent of this equation is divergent, as discussed in section \ref{subsec:electron_SE_im} [see discussion before (\ref{eqn:electronSelfEnergyImPartLowTemperature})]. The limit (\ref{eqn:momentumRelaxationTimeNoScreening}) is coincident  with the $T^{3}$ dependence found in Ref.~\cite{Zhang2013}, where vertex screening in the particular case of GaAs is not taken into account.
The neglect of vertex screening is acceptable in the temperature regime
$2r_s T_{\text{BG}} \lesssim T \ll T_{\text{BG}}$ in those materiales with $4r_s \ll 1$, because in that case the integral in Eq. (\ref{eqn:momentumRelaxationTime}) is dominated by exchanged momenta $q$ such that $k_{\rm TF} \ll q \ll k_F$, which are little sensitive to vertex screening. 
%By contrast, at temperatures lower than $r_s T_{\text{BG}}$, the important scattering processes are vertex-screened and the damping rate follows the $T^{5}$ law shown in Eq. (\ref{eqn:tau-trans-T5-law}).  
This intermediate regime of temperatures does not exist for substrate materials such that $2r_s \sim 1$.

%A similar situation is found for the electron decay caused by intrinsic deformation phonons: at these low temperatures, the electron inverse momentum relaxation time due to intrinsic phonons may decay like $T^4$ or $T^6$ depending on the details of the DFT calculation \cite{Kaasbjerg2012}, which can be interpreted as an effect of the screened interaction at low $q$ processes. \todo{[[This statement must be fully confirmed after a more detailed reading of Ref. \cite{Kaasbjerg2012}. We can only make this point is if we are sure that the difference between the two temperature dependence is caused by something very similar to our vertex screening. If we are not sure, then we must leave out this comment and perhaps find another excuse to cite Ref. \cite{Kaasbjerg2012}]]}

For the high temperature range $T \gg T_{\text{BG}}$, we have
\begin{equation}
\frac{\hbar}{2\,\tau_{+\text{tr}}^{\text{PA}}(k_{F})} \simeq \frac{\lambda}{2}
k_B T \int\limits_{0}^{1} x \sqrt{1-x^{2}} \,dx = \frac{\lambda}{6}k_{B}T~,
\label{eqn:momentumRelaxationTime-high-T}
\end{equation}
to be compared with Eq.~(\ref{eqn:electronSelfEnergyImPartHighTemperature}). The absence of a qualitative change in the temperature dependence as we shift from non-transport to transport lifetime is due to the relatively small weight, at high temperatures, of the transport-reduced, low-$q$ processes.

Thus we see that the transport scattering rates are comparable to the previous imaginary self-energies
except for an extra $(T/T_{\text{BG}})^{2}$ factor appearing at low
temperatures due to extra angular suppression of the otherwise dominant low-$q$ events.
A similar comparison holds for the intrinsic acoustic
deformation-potential phonons, where
\begin{equation}
\frac{\hbar}{2}\tau_{+\text{tr}}^{\text{DA}}
(k_{F})^{-1} \simeq 10 \left(\frac{T}{T_{\text{BG}}}\right)^{2} \;
\text{Im}\, \Sigma_{+}^{\text{DA}}(k_{F},0) \label{tr-nontr-low-T}
\end{equation}
at low temperatures, while
\begin{equation}
\frac{\hbar}{2}\tau_{+\text{tr}}^{\text{DA}} (k_{F})^{-1} \simeq
\frac{1}{2} \; \text{Im}\, \Sigma_{+}^{\text{DA}}(k_{F},0) \label{tr-nontr-high-T}
\end{equation}
for high temperatures. In the last two equations
%Eqs. (\ref{tr-nontr-low-T}) and (\ref{tr-nontr-high-T}) 
we are comparing the results of Refs. \cite{Zhang2013} and \cite{Li2013} for the transport scattering rate and the inverse lifetime, respectively.

In analogy with section \ref{subsec:electron_SE_im}, we may compare the transport rates due to deformation and piezoelectric phononic modes. In the low temperature limit (as before, $[k_F]$ is $k_F$ in units of $10^6\,\text{cm}^{-1}$),
\begin{equation}
\frac{\tau_{+\text{tr}}^{\text{PA}} (k_{F})^{-1}}{\tau_{+\text{tr}}^{\text{DA}} (k_{F})^{-1}} \simeq \frac{200}{[k_F]^{2}} \lambda \overline{\varepsilon}_{0}^{2} \frac{k_{B}T}{1\; \text{meV}}~,
\label{eqn:ratioMomentumRelaxationTimePADA-lowT}
\end{equation}
while at temperatures above $T_{\rm BG}$,
%(now there is no need for the previous function $f(x)$ linked to the logarithmic divergence):
\begin{equation}
\frac{\tau_{+\text{tr}}^{\text{PA}} (k_{F})^{-1}}{\tau_{+\text{tr}}^{\text{DA}} (k_{F})^{-1}} \simeq \frac{45}{[k_F]} \lambda~, 
\label{eqn:ratioMomentumRelaxationTimePADA}
\end{equation}
independent of temperature. Upon inserting the specific material parameters, Eq. (\ref{eqn:ratioMomentumRelaxationTimePADA}) is in agreement with the calculations of Ref.~\cite{Zhang2013}, where PA and DA transport rates are compared for GaAs.
Equations (\ref{eqn:ratioMomentumRelaxationTimePADA-lowT}) and (\ref{eqn:ratioMomentumRelaxationTimePADA}) must be compared to Eqs. (\ref{eqn:ratioElectronSelfEnergyImPartPADA-lowT}) and (\ref{eqn:ratioElectronSelfEnergyImPartPADA}) of section \ref{subsec:electron_SE_im}, respectively. Like in the non-transport lifetime estimates there presented, we note that piezoelectric dominate over deformation at non-small couplings and low densities.
We recall that Ref. \cite{Zhang2013} 
used a deformation constant $D=6.8$ eV, quite smaller than the value $D=25\, \text{eV}$ \cite{Li2013} we have used here. That replacement reduces $1/\tau^{\text{DA}}$ by about a factor of ten and makes  the substrate PA phonons relatively more important.

%We also note that the quotient in Eq.~(\ref{eqn:ratioMomentumRelaxationTimePADA}) is comparable or usually lower than that of Eq.~(\ref{eqn:ratioElectronSelfEnergyImPartPADA}).
%\todo{[[lo que viene a continuación está todavía sujeto a revisión]]}
%The graphene intrinsic acoustic phonons can dominate the inverse mobility for low piezoelectricities with $\lambda \lesssim 0.01$) or high carrier concentrations. At densities higher than $\sim 5\times 10^{11} \, \text{cm}^{-2}$ the DA scattering dominates PA for GaAs, but for materials with $\lambda$ close to 0.1, this critical density exceeds $10^{13} \, \text{cm}^{-2}$.
%%, if the deformation constant $\Xi$ is used, or above $10^{13}\, \text{cm}^{-2}$ taking $D$.

Finally, in order to compute the electron mobility we average the momentum relaxation time [see Eq. (\ref{eqn:momentumRelaxationTime})],
\begin{equation}
\overline{\tau_{\text{tr}}} \equiv \int d\epsilon D(\epsilon)\tau_{+\text{tr}}(k(\epsilon)) [-dn_{F}(\epsilon)/d\epsilon] \, ,
\end{equation}
and because the energy derivative peaks at $E_{F}$ while $\tau_{+\text{tr}}(k)$ varies slowly with $k$, one can write the classical Drude formula for the mobility, 
\begin{equation}
\mu = \frac{e\,\tau_{+\text{tr}}(k_{F})}{m^{*}} 
%= \frac{e v_{F}}{\hbar k_{F}} \tau_{\text{tr}}(k_{F})
~,
\label{eqn:mu}
\end{equation}
in terms of $\tau_{+\text{tr}}(k)$ computed at the Fermi level and the ``effective mass'' $m^*=\hbar k_F / v_F$ of the graphene Dirac fermions.

%%%%%%%%%%%%%%%%%%%%%%%%%%%%%%%%%%%%%%%%%%%%%%%%%%%%%%%%%%%%%%%%%%%%%%%%%%%%%%%%%%%%%%%%%%%%%%%%%%%%
\section{Numerical results}
\label{sec:numerical}
%%%%%%%%%%%%%%%%%%%%%%%%%%%%%%%%%%%%%%%%%%%%%%%%%%%%%%%%%%%%%%%%%%%%%%%%%%%%%%%%%%%%%%%%%%%%%%%%%%%%
In the following, we present and discuss numerical results for the various rates and mean free paths derived in sections
\ref{subsec:electron_SE_im} and \ref{subsec:mobilities}.
Unless otherwise stated, the numerical values of this section are computed
for ZnO substrates (Z-cut), which is isotropic (see Fig.
\ref{fig:vertexPlot})
and whose parameters are
$\lambda = 0.03$ and $\overline{\varepsilon}_{0}=4.8$, which implies $k_{\rm TF}/k_F \simeq 2$ and $k_B T_{\rm BG}/E_F \simeq 0.0054$.

\begin{figure}
	\raggedleft
	\includegraphics[width=0.95\linewidth]{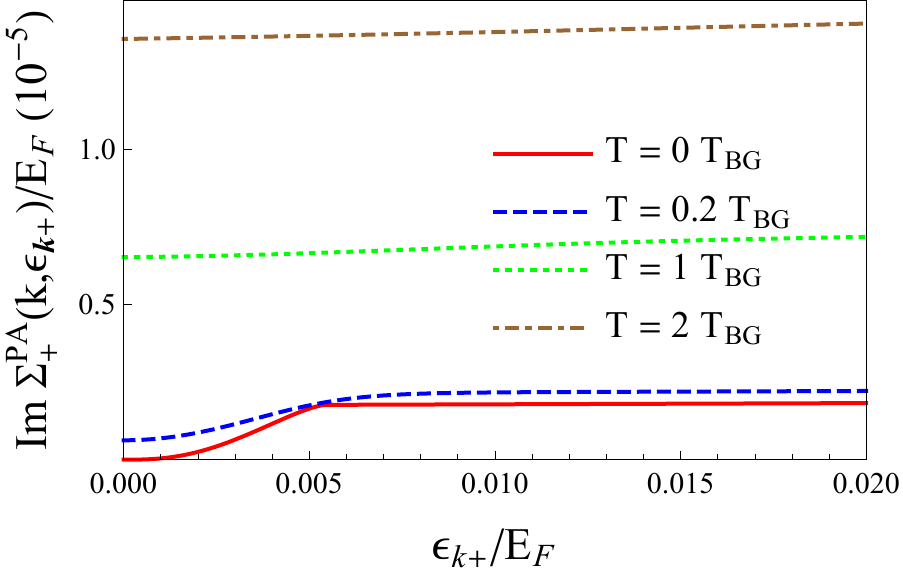}
	\vskip 0.2cm
	\includegraphics[width=0.94\linewidth]{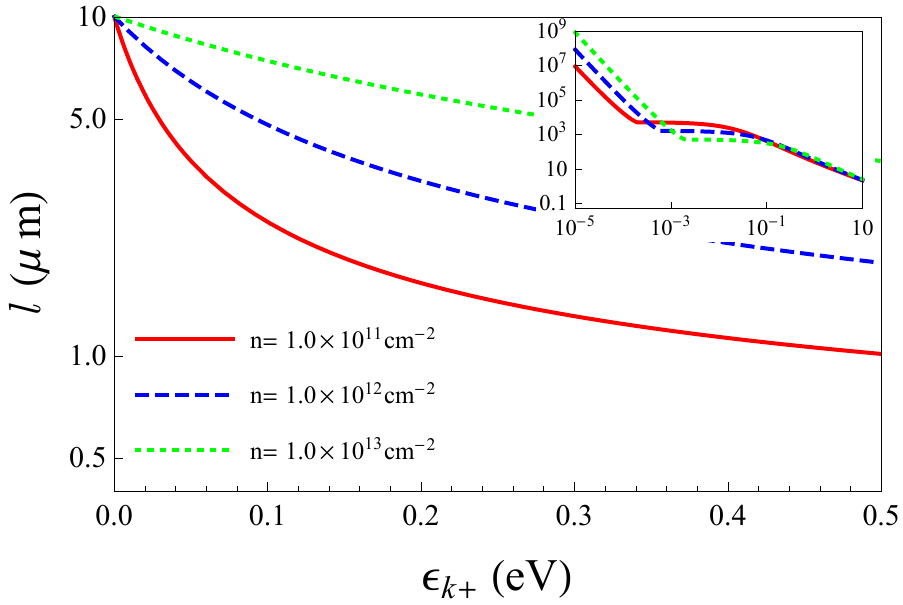}
	\caption[comentarios breves]{(Color online) Imaginary part of the self-energy and inelastic scattering length of charge carriers in graphene on ZnO (Z-cut) as a function of the energy,
		${\epsilon_{k}}/{E_{F}} = (k-k_{F})/{k_{F}}$.
		Upper: $ \text{Im}\, \Sigma_{+}^{\text{PA}}$ for different temperatures.
		The curves are valid for all densities.
		Lower: Inelastic mean free path $l$ for different carrier concentrations at room temperature ($T=300$ K $=26$ 
		meV/$k_B$). The inset shows $l$ at $T=0$ for the same densities.
		The Bloch-Gr\"uneisen temperature $T_{\rm BG}$ is given in Eq.~(\ref{eqn:TBG_def}).
		For this material, $k_B T_{\rm BG}=0.0054 E_F$ (for all carrier densities) and $k_{\rm TF}/k_{\rm F}\simeq 2$ (thus
		$T_{\rm BG} \simeq \hbar v_s k_{\rm TF}/k_B$).
		For these three densities, $k_B T_{\text{BG}}=0.2, 0.63, 2$ meV, while  $E_{F}=37.4, 117, 374$ meV.
	}
	\label{fig:imSE_MFPvsQPEnergy}
\end{figure}

In upper Fig. \ref{fig:imSE_MFPvsQPEnergy}, we show
%based on the ``universal'' function for the adimensionally scaled
the imaginary part of the on-shell self energy  as a function of the  parameter
$\epsilon_{k}/E_{F}>0$
for different temperatures.
The curves are universal in the sense of density independent.
The zero temperature curve shows, for small $\epsilon_{k}$, the
limiting $\epsilon_{k}^3$ behavior of  Eq.~(\ref{eqn:electronSelfEnergyImPartZeroTemperature}),
which arises due to the combined effect of screening and the phase space restrictions faced by
the electrons when losing energy via phonon emission. 
This restriction disappears
when $\epsilon_{k}$ is greater than any phononic
energy, i.e.,  $\epsilon_{k} \gg k_{B} T_{\text{BG}}$.
Above this threshold the imaginary part of the self-energy becomes energy independent, as predicted by Eq. (\ref{eqn:hvskTF-eps-EF}). At still higher energies ($\epsilon_{k} \gg E_F$, not shown in upper Fig. \ref{fig:imSE_MFPvsQPEnergy}), it increases linearly with the
length of the constant energy circumference at the quasiparticle energy
$E_{k+}\propto k$. Such a linear increase with $k$ would appear with a negligible slope in the tiny scale of $\epsilon_{k} \propto (k-k_F)$ of upper Fig. \ref{fig:imSE_MFPvsQPEnergy}. Specifically, the slope is, in the dimensionless units of upper Fig. \ref{fig:imSE_MFPvsQPEnergy}, $(\lambda\pi/16)(v_s/v_F)$.

Upper Fig. \ref{fig:imSE_MFPvsQPEnergy} also shows that a further increase in temperature ($T>T_{\text{BG}})$ smears these features due to phonon excitation and electron heating near the Fermi energy, as exemplified in Eq. (\ref{eqn:electronSelfEnergyImPartHighTemperature}).

\begin{figure}
	\raggedleft
	\includegraphics[width=0.93\linewidth]{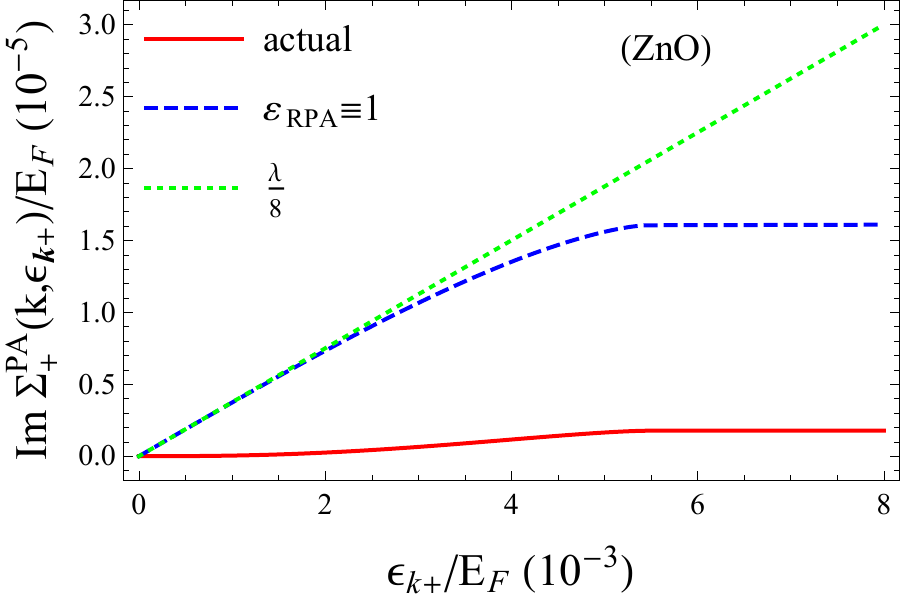}
	\vskip 0.5cm
	\includegraphics[width=0.9\linewidth]{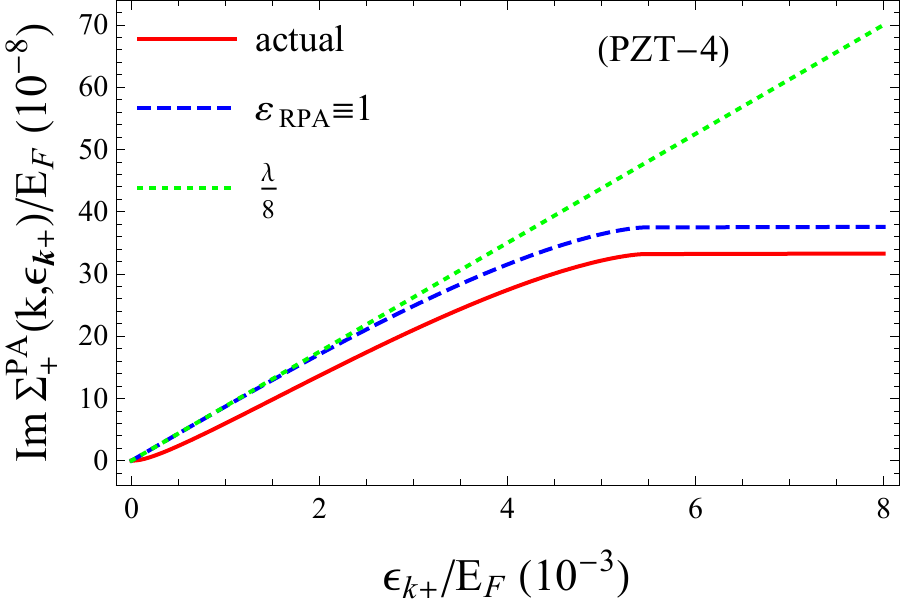}
	\caption[comentarios breves]{(Color online) Low-energy behavior of the imaginary part of the
		self-energy at zero temperature.
		The solid, dashed
		and dotted lines correspond, respectively, to the exact values, the values without vertex screening, and the values obtained (in the unscreened case) from the linear $\lambda/8$ approximation of  Eq.~(\ref{eqn:electronSelfEnergyImPartZeroTemperatureNoScreening}).
		Upper: Graphene on ZnO (Z-cut). Lower: Graphene on PZT-4 (Z-cut), for which $k_B T_{\rm BG}/E_F=0.0045$, $k_{\rm TF}/k_{\rm F}\simeq 0.025$, and $\hbar v_s k_{\rm TF}/E_F \simeq 5.7 \times 10^{-5}$. See Table \ref{tab:materials} for $\lambda$ values.
	}
	\label{fig:imSE_vsQPEnergy_ZnOPZT}
\end{figure}

The effect of vertex screening in the regime of low $\epsilon_{k}$, low $T$ can be appreciated in Fig.~\ref{fig:imSE_vsQPEnergy_ZnOPZT},
for both ZnO and (angle averaged) PZT substrates with its higher dielectric constant (and thus smaller $\lambda$).
For the sake
of comparison, the graphics include also the linear approximation
(\ref{eqn:electronSelfEnergyImPartZeroTemperatureNoScreening}), which holds better for PZT because its large dielectric constant reduces the size of the phase space region where the screening of the phonon interaction
by the electron cloud (vertex correction) is really important. Unlike for ZnO, in this material $k_{\text{TF}}$ is considerably smaller than $k_F$, which leaves room for an intermediate range of $\epsilon_k$ values for which the approximation $\varepsilon_{\rm RPA} \simeq 1$ is acceptable while the linear behavior still holds. As announced in section \ref{subsec:electron_SE_im}, after Eq. (\ref{eqn:electronSelfEnergyImPartZeroTemperatureNoScreening}), there is an offset between the true imaginary self-energy and the linear approximation due to the reduced contribution of the screened low-$q$ processes.

\begin{figure}
	\raggedleft
	\includegraphics[width=0.97\linewidth]{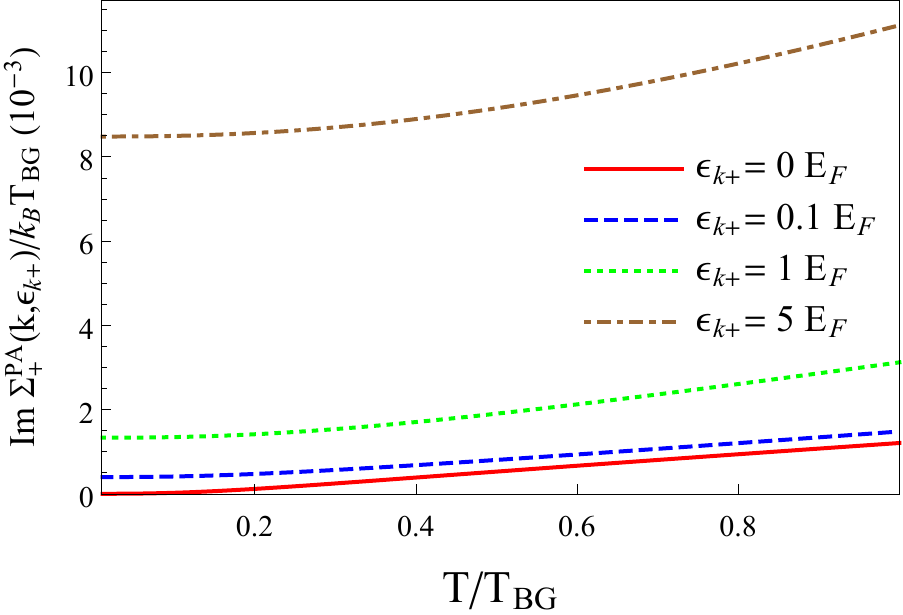}
	\vskip 0.2cm
	\includegraphics[width=\linewidth]{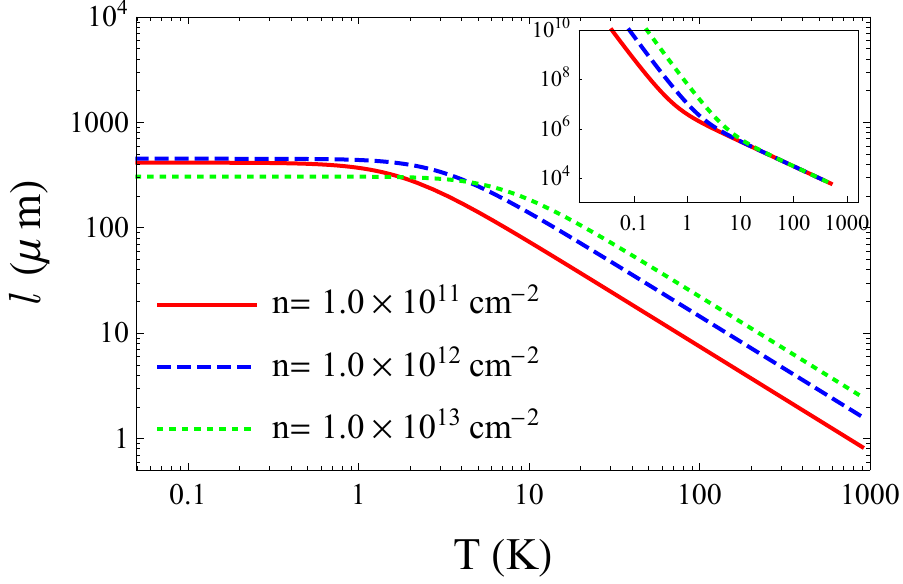}
	\caption[comentarios breves]{(Color online) Imaginary part of the self-energy
		and inelastic scattering length. Upper: $ \text{Im}\,
		\Sigma_{+}^{\text{PA}}$ as a function of $T/T_{\text{BG}}$ for
		different values of $\epsilon_{k}/E_{F}$. Lower: $l$ as a
		function of $T$ for $\epsilon_{k}=0.1\,\text{eV}\simeq 1160 \, \text{K}$ and different
		doping levels. The inset shows the corresponding curves for $\epsilon_{k}=0$. Here, $E_F\simeq 185\, k_B T_{\rm BG}$.
		The values of $T_{\text{BG}}$ for these three densities are $2.35,7.35,23.5$ K.
		}
	\label{fig:imSE_MFPvsQPEnergyVariousTemperatures}
\end{figure}

Upper Fig. \ref{fig:imSE_MFPvsQPEnergyVariousTemperatures} shows the temperature
dependence of
$\text{Im}\, \Sigma_{+}^{\text{PA}}$ for fixed values of $\epsilon_k$.
%effects can be seen to be more important than these induced from the distance
%of the quasiparticle state to the Fermi level. Of course,
At low temperatures ($T \ll \epsilon_{k}$, hot electron regime),
these decay linewidths are independent of $T$. 
Note that in this figure the nonzero values of $\epsilon_k$ are well above $\hbar v_s k_{\rm TF}$ and thus the limit (\ref{eqn:electronSelfEnergyImPartZeroTemperature}) does not apply.
At higher temperatures ($T>T_{\text{BG}}$),
the linear behavior of Eq.~(\ref{eqn:electronSelfEnergyImPartHighTemperature}) is recovered.

The lower Figs. \ref{fig:imSE_MFPvsQPEnergy} and \ref{fig:imSE_MFPvsQPEnergyVariousTemperatures} are devoted to the inelastic scattering mean free path, which is the inverse of the imaginary part of the on-shell
self-energy:
\begin{align}
l(k)
&= \frac{\hbar
	v_{F}}{2\,\text{Im}\,
	\Sigma_{+}^{\text{PA}}(k,\epsilon_{k})}
,
\label{eqn:l_def}
\end{align}
Lower Fig. \ref{fig:imSE_MFPvsQPEnergy} shows
values for $l(k)$
as a function of $\epsilon_{k}$ for three cases of typical doping
conditions. Note that they tend to coincide at small $\epsilon_{k}$,
as suggested by Eq. (\ref{eqn:electronSelfEnergyImPartHighTemperature}) (case $T>T_{\rm BG}$), which predicts a doping-independent low-$\epsilon_k$ ($k\rightarrow k_F$) limit at nonzero temperatures. Finally, the inset of lower Fig. \ref{fig:imSE_MFPvsQPEnergy} clearly displays the three energy regimes that hold at zero temperature and which can be inferred from Eqs. (\ref{eqn:electronSelfEnergyImPartZeroTemperature})-(\ref{eqn:hvskTF-EF-eps}).

The temperature dependence of $l$ is shown in the lower Fig. \ref{fig:imSE_MFPvsQPEnergyVariousTemperatures}. A crossover from ($T$-independent) low-temperature to ($T^{-1}$) high-temperature behavior can be appreciated for $T \sim T_{\text{BG}}$, in agreement with Eqs. (\ref{eqn:electronSelfEnergyImPartZeroTemperature}) and (\ref{eqn:electronSelfEnergyImPartHighTemperature}). One must note however that Eq. (\ref{eqn:electronSelfEnergyImPartZeroTemperature}) does not truly apply to the low-temperature sector of this graph because here $\epsilon_k > \hbar v_s k_{\rm TF}$, unlike assumed in (\ref{eqn:electronSelfEnergyImPartZeroTemperature}). This explains the discrepancy in the density dependence. For this material, $\hbar v_s k_{\rm TF}$ takes values 0.2, 0.63, 2 meV for the three listed densities, all much smaller than the value $\epsilon_k=100$ meV there considered.

The inset shows the corresponding curves for $\epsilon_k=0$. A clear crossover for $T^{-3}$ to $T^{-1}$ behavior is observed at $T \sim T_{\text{BG}}$, in agreement with Eqs. (\ref{eqn:electronSelfEnergyImPartLowTemperature}) and (\ref{eqn:electronSelfEnergyImPartHighTemperature}).

\begin{figure}[b]
	\raggedleft
	\includegraphics[width=\linewidth]{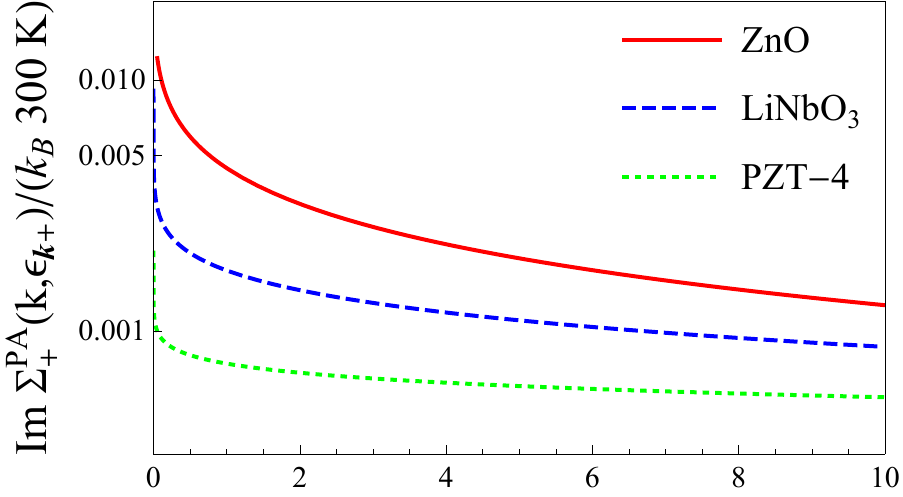}
	\vskip 0.5cm
	\includegraphics[width=0.96\linewidth]{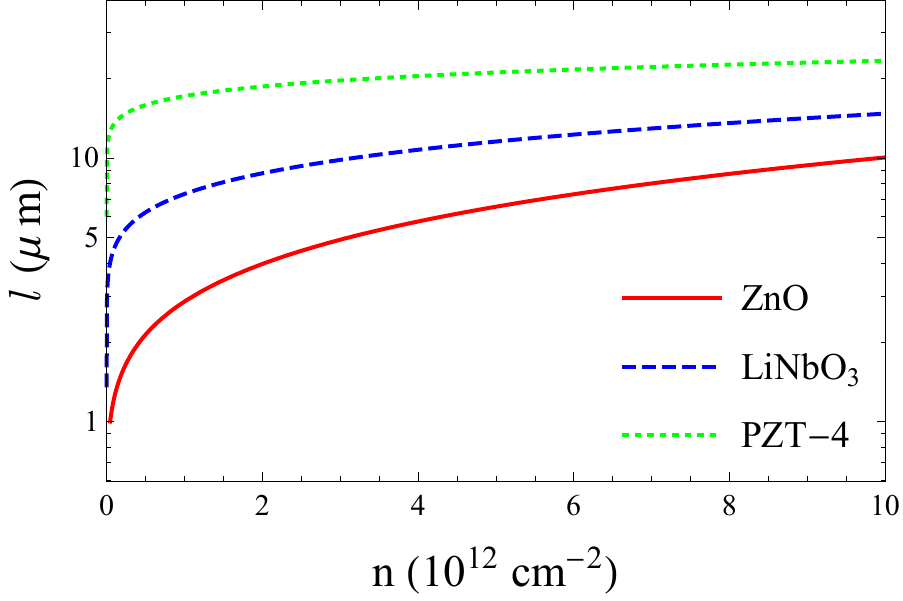}
	\caption[comentarios breves]{(Color online) Imaginary part of the self-energy and inelastic scattering length as a function of the doping, for different materials,
		at fixed (room) temperature and electronic state $k = \sqrt{\pi\times 10^{13}\,\text{cm}^{-2}}$ (recall $k_F=\sqrt{\pi n}$).
		Upper: $ \text{Im}\, \Sigma_{+}^{\text{PA}}$ as a function of  carrier density $n$. Lower: $l$ as a function of $n$ in the same units for the same materials.
	}
	\label{fig:imSE_MFPvsDoping}
\end{figure}

For a fixed value of $k$ and at room temperature,
%$k=\sqrt{\pi \times 10^{13}\,\text{cm}^{-2}}$ and $E_{k+}=0.369\,\text{eV}$,
Fig.  \ref{fig:imSE_MFPvsDoping} shows the variation of $\text{Im}\,
\Sigma_{+}^{\text{PA}}$ and of the mean free path
as a function of the carrier density.
%(a) shows the change of $\frac{ \text{Im}\,
%\Sigma_{+}^{\text{PA}}(\mathbf{k},\epsilon_{k}) }{k_{B}T}$ at fixed
%room temperature $T=300\,K$ when the Fermi level is varied till the limit
%$E_{k_{F}} = E_{k+}$ at $n=10^{13}\,\text{cm}^{-2}$ and (b) displays the
%associated mean free path.
A logarithmic divergence in the linewidth, accompanied by a vanishing mean free path,
is seen to appear
in the undoped regime,  where the description of the system employed in the present paper is not valid anymore. This spurious low-doping behavior can be expected from an extrapolation of Eq. (\ref{eqn:electronSelfEnergyImPartHighTemperature}) to low doping.

\begin{figure}
	\raggedleft
	\includegraphics[width=\columnwidth]{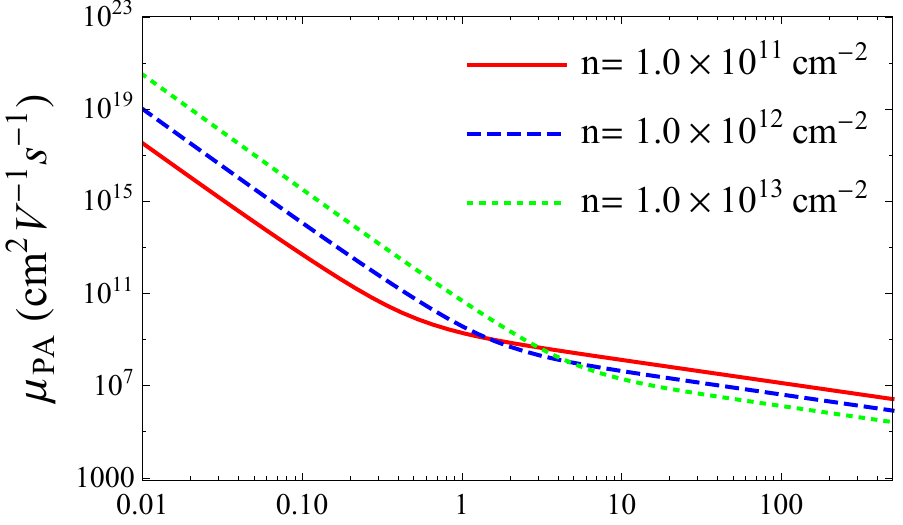}
	\vskip 0.5cm
	\includegraphics[width=0.96\columnwidth]{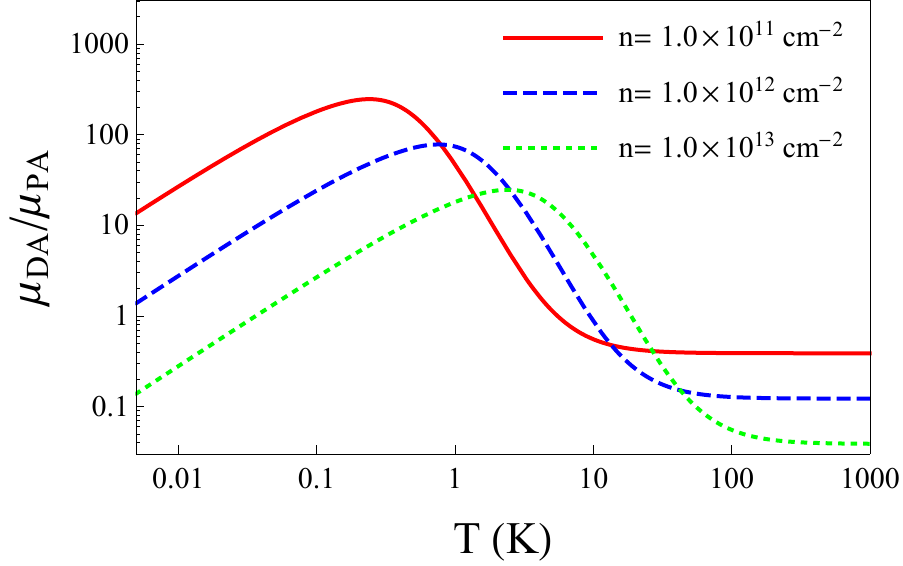}
	\caption[comentarios breves]{(Color online) Electron mobility due only to piezoelectric acoustic phonons and its comparison with that due to intrinsic phonons. Upper: The mobility $\mu_{\text{PA}}$ as a function of the temperature, for several carrier concentrations.
		Lower: The ratio $\mu_{\text{DA}}/\mu_{\text{PA}}$, where $\mu_{\text{DA}}$ is the mobility obtained when only the deformation potential of intrinsic phonons (with $D=25\,\text{eV}$) is included. The ratio between mobilities must be increased by a factor $(25/6.8)^2\simeq 13.5$ when the value $D=6.8\,\text{eV}$ is used \cite{Zhang2013}. The values of $T_{\text{BG}}$ for these doping levels are given in the caption of Fig. \ref{fig:imSE_MFPvsQPEnergyVariousTemperatures}, while $T_F=434,1360,4340$ K.
	}
	\label{fig:mobilityVsTemperature}
\end{figure}

In the upper Fig.~\ref{fig:mobilityVsTemperature}, we show the electron mobility $\mu$ [see Eq.~(\ref{eqn:mu})] due only to piezoelectric phonons.
The $T^{-5}$ and $T^{-1}$ behaviors can be appreciated at low and high temperatures, respectively, as expected from Eqs. (\ref{eqn:tau-trans-T5-law}) and (\ref{eqn:momentumRelaxationTime-high-T}) taking into account Eq. (\ref{eqn:mu}) for the density dependence.

Finally, in the lower Fig.~\ref{fig:mobilityVsTemperature} we compare the substrate induced mobility to that stemming only from graphene intrinsic phonons, with $D=25\,\text{eV}$. The total combined mobility due to (piezoelectric and intrinsic deformation) acoustical phonons is $\mu = \left(\mu_{\text{PA}}^{-1} + \mu_{\text{DA}}^{-1}\right)^{-1}$. Specifically, we plot the ratio between the two inverse mobilities. The smaller value of $D=6.8\,\text{eV}$ reduces the intrinsic inverse mobility by an order of magnitude and correspondingly increases the relative importance of piezoelectric phonons. This ratio between transport scattering rates shows two clear low- and high-$T$ regimes with linear-in-$T$ and $T$-independent behaviors, respectively, in agreement with  Eqs. (\ref{eqn:ratioMomentumRelaxationTimePADA-lowT}) and (\ref{eqn:ratioMomentumRelaxationTimePADA}).
At low and high temperatures, the relative importance of the PA phonons increases with decreasing density. There is an intermediate temperature regime in which the density dependence is inverted.
Thus we see that the piezoelectric phonons dominate over a wide range of temperatures and densities. If $D=6.8\,\text{eV}$ for the intrinsic phonons is chosen, then the momentum relaxation due to PA phonons here computed prevails essentially always except at very high temperature and density or for extremely low temperatures.

\section{Conclusions}
We have studied the effective interaction of charge carriers in graphene on a piezoelectric substrate,
as modified by the acoustic phonons of the piezoelectric substrate.
Our diagrammatic approach takes into account the renormalization of both phonon modes and carrier states
due to the mutual interaction,
and emphasizes the importance of all the involved screening processes for a correct evaluation of
the mean free path and carrier mobility.
We have obtained numerous analytical limits as a function of carrier energy, density and temperature, which have allowed us to understand the trends shown by the numerical results.

When compared with the values obtained when only intrinsic deformation phonons are taken into account, we find that the contributions of the piezoelectric acoustic phonons to the inverse lifetime and mobility dominate over a considerable range of temperatures and doping levels, a parameter range that becomes almost pervasive if low values of the deformation coupling constant are chosen from the literature.

As our results are applicable to piezoelectric materials of various lattice symmetries and interaction strengths,
they will be helpful in the development of electronic devices involving graphene deposited on piezoelectric
substrates.

\begin{acknowledgments}
We wish to thank Fernando Calle and Jorge Pedr\'os for valuable discussions.
This work has been supported
by the Spain's MINECO
through Grants No. FIS2011-23713, FIS2013-41716-P;
the European Research Council Advanced Grant (contract
290846), and the European Commission under the
Graphene Flagship, contract CNECTICT- 604391.
DGG acknowledges financial
support from Campus de Excelencia Internacional (Campus
Moncloa UCM-UPM).
\end{acknowledgments}

\bibliographystyle{apsrev4-1}
\bibliography{pSAWManyBody}

\end{document}